\documentclass[aps,pre,reprint,twocolumn,amsmath,amssymb,groupedaddress]{revtex4-1}
\usepackage{graphicx}
\usepackage[hidelinks]{hyperref}
\usepackage[]{units}

\usepackage{verbatim}
\usepackage{color}
\begin{document}

\title{Dynamics of particle uptake at cell membranes}

\author{Felix Frey}
\author{Falko Ziebert}
\author{Ulrich S. Schwarz}

\affiliation{Institute for Theoretical Physics, Heidelberg University, Philosophenweg 19, 69120 Heidelberg, Germany 
and BioQuant, Heidelberg University, Im Neuenheimer Feld 267, 69120 Heidelberg, Germany}

\date{\today}

\begin{abstract}
Receptor-mediated endocytosis requires that the energy of adhesion 
overcomes the deformation energy of the plasma membrane. 
The resulting driving force is balanced by dissipative forces,
leading to deterministic dynamical equations. While the shape
of the free membrane does not play an important role for 
tensed and loose membranes, in the intermediate regime
it leads to an important energy barrier. Here we show that this barrier is
similar to but different from an effective line tension and suggest
a simple analytical approximation for it. 
We then explore the rich dynamics of uptake for particles of different shapes
and present the corresponding dynamical state diagrams.
We also extend our model to include stochastic fluctuations, which
facilitate uptake and lead to corresponding changes in the phase diagrams.
\end{abstract}

\maketitle

\section{Introduction}

The plasma membrane presents a physical barrier that separates the interior of the cell
from its environment. Therefore, the ability of cells to exchange information and material across their
plasma membrane is of central importance for their function \cite{auth2018interaction,nano_up}.
On the one hand, these uptake processes are vital for nutrient influx and signal transduction \cite{alberts}.
On the other hand, pathogens like viruses hijack cellular uptake 
processes to enter host cells during infections \cite{kumberger2016}. 
In addition, uptake of artificial particles at cell membranes can be desired, as e.g. in the 
context of drug delivery \cite{drugdelivery}, or undesired, as e.g. in the 
context of microplastics \cite{Moos}.

\begin{figure}[t]
\centering
\includegraphics[ width=0.48 \textwidth]{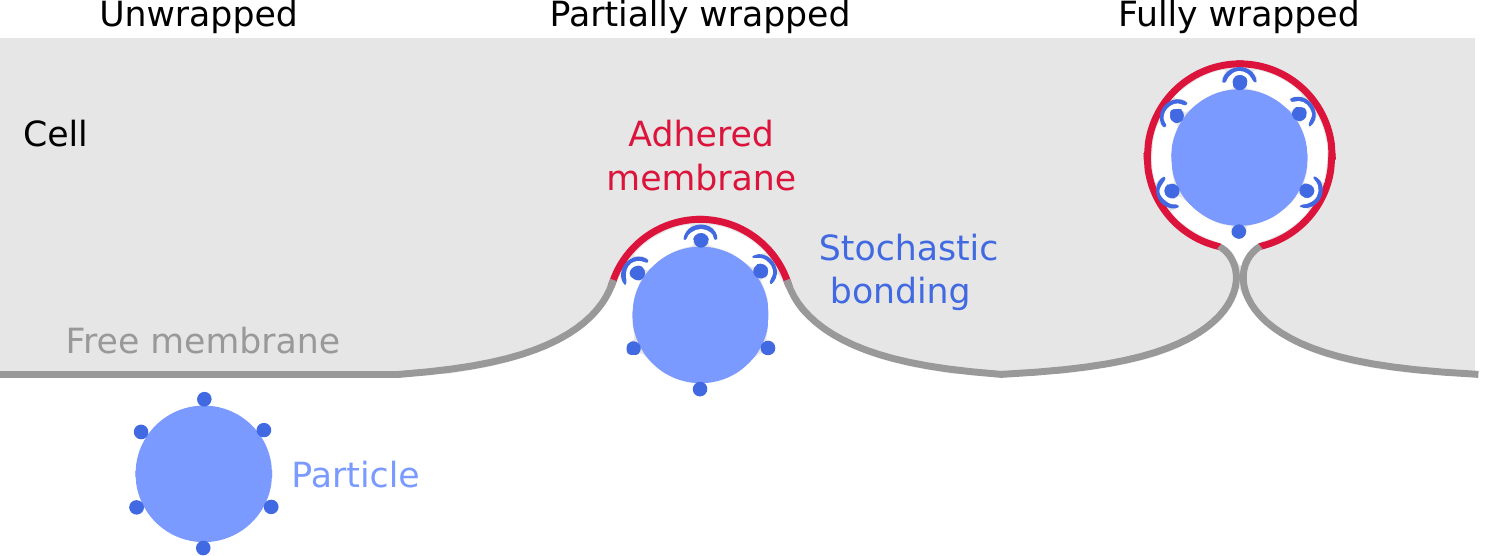}
\caption{During receptor-mediated endocytosis, the particle
goes from unwrapped through partially wrapped to fully wrapped. 
While membrane shape is fixed by particle shape in the region of the
adhered membrane (red part of the contour), the shape of the
free membrane follows from an energy minimization (grey part of the contour).
}
\label{fig:Figure1}
\end{figure}

In receptor-mediated endocytosis particles with sizes between
$\unit[10-300] {nm}$ are taken up because the energy gain upon particle
binding to cell surface receptors overcomes the deformation 
energy of the membrane \cite{gao2005}.
Membrane shape is of central importance for this process.
It is fixed by particle shape at the adhered part, but follows
from minimization of the membrane energy for the free part,
compare Fig.~\ref{fig:Figure1}. Very importantly, 
cargo particles can come with a huge diversity in shape,
including the case of viruses. The most frequent virus shape
is the spherical or icosahedral shape, followed by filamentous and then by
more complex shapes. To name a few examples, reovirus, causing respiratory or 
gastrointestinal illnesses,
has icosahedral shape \cite{barton2001}, Marburg or Ebola viruses
have filamentous shape \cite{Ebola2}, and rabies virus has
bullet-like shape \cite{matsumoto1963electron}. Apart from shape,
stochastic fluctuations in receptor binding might also play a role, as the 
cargo particles are small and typically covered by only few tens of ligands.

The uptake of small particles has been previously studied both 
analytically and by computer simulations.
Deterministic approaches usually focus on calculating minimal
energy shapes for the plasma membrane and the attached particle
to deduce dynamical state diagrams 
\cite{LipoDoeb,DesernoPRE04,agudo2015},
investigate uptake dynamics and the role of receptor diffusion within the 
plasma membrane  \cite{gao2005,decuzzi2007role}, study the consequences of 
particle elasticity during uptake \cite{yi2011cellular,zeng2017contact} or 
interactions of the particle and the cytoskeleton \cite{sun2006}.
Stochastic approaches usually focus on the effect of ligand-receptor
binding \cite{yi2017kinetics}.
Computer simulations complement these studies by considering 
especially the role of non-spherical particle geometries
\cite{vacha2011receptor,huang2013role,dasgupta2014,Bahrami,deng2019entry}
or the role of scission, when the wrapped particle is separated from the membrane
\cite{mcdargh2016constriction}.
  
Despite this plethora of different approaches, analytical approaches are
rare that allow us to study the interplay of particle shape, free membrane
shape and stochasticity in one transparent framework. 
Recently we showed that in a deterministic model,
spherical particles are taken up slower compared to cylindrical particles, whereas
the situation can reverse in a stochastic description, because spherical particles
profit more from the presence of noise \cite{frey2019}. However,
in this earlier work we have neglected the exact role of the free part of the membrane
and did not investigate the possibility that the dynamics stops with a partially wrapped particle (Fig.~\ref{fig:Figure1}).
Earlier it has been suggested that the free membrane might act 
as an effective line tension \cite{LipoDoeb} and exact formulae for the
elastic energy barrier provided by the free membrane have been
derived for the limits of tense and loose membranes \cite{foret2014}. Here we present
a comprehensive study of these important effects and show that
in the general case, a simple term that is similar to but different from 
a line tension term describes this energy barrier well. With this
simplification, we are able to perform a comprehensive analysis 
of the uptake dynamics at membranes, including the effect of particle
shape and stochastic fluctuations. By combining models
for membrane mechanics, overdamped membrane dynamics and stochastic 
dynamics of receptor-ligand binding, we find a very rich scenario of possible uptake dynamics.

Our work is organized as follows.
In section~\ref{sec:two} we first calculate the shape and energy of the 
free membrane by numerically solving the shape equations that give the minimal 
energy shapes of the free membrane. By comparing this energy to the energy
of the adhered membrane we can identify the parameter regime in which the free membrane
cannot be neglected, as it contributes up to $20\%$ of the total energy.
We also show that this energy contribution is similar to but different from the
effect of a line tension; in particular, it leads to faster initial uptake
and the associated energy barrier is located at higher coverage. 
These effects can be described well by a simple analytical ansatz introduced here.
In section~\ref{sec:three} we derive and analyze the deterministic dynamical equations
for cylindrical, spherical and spherocylindrical particles. Here the free membrane
is not yet taken into account explicitly, but its potential effect can already be appreciated
on the level of a line tension.
We find that spherical particles are taken up slower than cylindrical particles.
In addition, we find that short cylinders are taken up faster in normal orientation, whereas
long cylinders are taken up faster in parallel orientation.
For spherocylindrical particles we find that they are taken up fastest in parallel 
orientation.
In section~\ref{sec:four} we include the analytically exact energy of the 
free membrane for parallel cylindrical particles into our dynamical model 
and calculate dynamical state diagrams. 
In addition, we also include free membrane effects by the phenomenological
approach. As the phenomenological approach gives similar results, we then apply our 
approach also to spherical particles.
In section~\ref{sec:five} we calculate dynamical state diagrams for
spherical particles using the phenomenological
ansatz. We identify three regimes: full uptake, partial uptake and no uptake
determined by membrane elasticity, adhesion energy and the free membrane.
In section~\ref{sec:six} we then extend our model to a stochastic description,
which in contrast to earlier work \cite{frey2019} now includes the effect of the free membrane. 
We find that fluctuations decrease uptake times and expand uptake
to parameters regions where uptake is not possible in the deterministic model.
In addition, we find that spherical particles can be taken up faster with fluctuations
compared to parallel cylindrical particles. 

\section{Membrane energies}
\label{sec:two}

To arrive at a dynamical model for particle uptake at cell membranes,
we follow the standard approach and first compare the energy gain due to adhesion 
and the energy cost due to bending and tension 
\cite{LipoDoeb,DesernoPRE04,agudo2015,auth2018interaction}.
Later we will introduce dynamics by also considering dissipative forces. 
We are interested in ligand-receptor interactions and for simplicity assume
that they are distributed homogeneously over the particle surface.
The total energy of the membrane is then described by the following
generalization of the Helfrich bending Hamiltonian \cite{helfrich1973} 
\begin{equation}
E=-\int_{A_\mathrm{ad}} W \mathrm{d} A + \int_{A_\mathrm{mem}}  
2\kappa H^2 \mathrm{d}A + \sigma \Delta A + \zeta \mathcal{E}\, .
\label{eq:free_energy_general}
\end{equation}
The first term is the gain in adhesion energy, where the adhesion energy
per area $W$ is defined to be positive. The second term is the bending energy
with $\kappa$ being the bending rigidity and $H$ the mean curvature of the membrane.
The third term is the tension energy, with $\sigma$ being the membrane tension 
and $\Delta A$ the excess area compared 
to the flat membrane. It is important to note that 
only the part $A_{\mathrm{ad}}$ of the membrane that adheres to 
the particle contributes to the adhesion energy,
while both the adhered and the free parts of the membrane, $A_{\mathrm{mem}}=A_{\mathrm{ad}}+A_{\mathrm{free}}$, 
contribute to bending and tension (compare Fig.~\ref{fig:Figure1}). 
The last term in Eq.~(\ref{eq:free_energy_general})
results from a possible line tension $\zeta$,
with $\mathcal{E}$ being the length of the edge between the membrane adhering to the particle 
and the free cell membrane. 
We note that in general Eq.~(\ref{eq:free_energy_general}) also includes
a term that depends on Gaussian curvature. However, since we do not consider 
topological changes, it can be neglected in our context \cite{DesernoPRE04}.

\begin{figure}[t]
\centering
\includegraphics[ width=0.48 \textwidth]{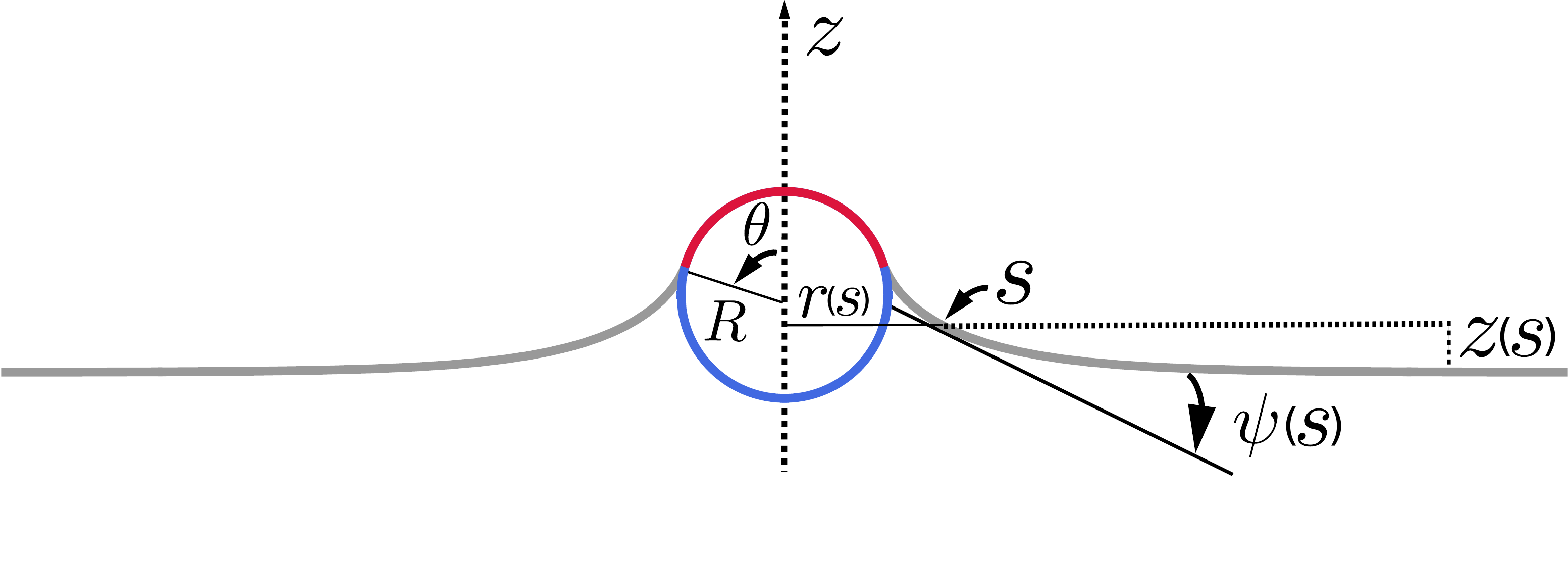}
\caption{Parametrization of membrane shape for a spherical particle. The adhered part of the
membrane is shown in red, the free part of the membrane is shown
in grey and the particle is shown in blue. $R$ is the particle radius, $\theta$ is the uptake angle
and $s$ the contour length of the free membrane. Tangent angle $\psi$, radial distance $r$ and height $z$
are functions of $s$.}
\label{fig:Figure2}
\end{figure}

The two membrane parameters $\kappa$ and $\sigma$ define a typical 
length scale $\lambda=\sqrt{\kappa / \sigma}$.
Using this scale, the membrane can be classified as tense ($\lambda/R\ll 1$) or
loose ($\lambda/R \gg 1$).
Considering typical parameter values occurring in the context
of particle uptake at cell membranes,
$\kappa=25\,{k_{\mathrm{B}}T}$ and $\sigma=10^{-5}-10^{-3}{\rm N/m}$ 
\cite{foret2014,kumar2016}, one has $\lambda=10-100\,{\rm nm}$. 
As typical sizes for virus or nanoparticles range from 
$R=10 - 100\,{\rm nm}$, it holds that $\lambda / R \sim 0.1-10$ and
the biologically most relevant regime is hence intermediate between tense and loose. 

The line tension term in Eq.~(\ref{eq:free_energy_general}) could result e.g.\
from the localization of certain lipids or proteins to the curved membrane
at the border between the adhered and free membrane \cite{lipowsky1992budding}. More importantly in our
context, however, such a term could potentially also be used to describe the effective
behaviour of the free membrane, even in the absence of a microscopic line tension \cite{LipoDoeb}. In this case, one could
restrict the integration of the bending energy over the adhered part of the membrane.
For dimensional reasons, one then would expect the effective line tension to scale
as $\zeta=\sqrt{\kappa \sigma}$ and the typical range
would be $\zeta=1-10\,{\rm pN}$ \cite{DesernoPRE04,foret2014}.

In order to discuss the role of the free membrane shape in more detail, let
us consider the uptake of a spherical particle.
Fig.~\ref{fig:Figure2} shows the used parametrization \cite{foret2014,foret2018mechanosensitivity}, 
where $\theta$ is the uptake angle, measured with respect to the symmetry axis (along the $z$-direction). 
The membrane contour is parameterized by its arc length $s$ 
relative to the point where the adhering membrane is connected to the free part. 
Furthermore, $\psi(s)$ is the angle between 
the radial axis normal to the $z$-axis and the contour tangent,
$r(s)$ is the radial distance to the $z$-axis, and $z(s)$ the height.
Note that $r$ and $z$ can be obtained by integration over $\psi$. 

For the spherical particle, the area adhering to the membrane 
as a function of particle radius $R$ and uptake angle $\theta$
is given by $A_{\mathrm{ad}}=2\pi R^2  (1-\cos \theta )$
and thus the adhesion energy would be $E^{\mathrm{W}} = - 2\pi WR^2 (1-\cos\theta)$.
Similarly, also the contributions from bending and tension of the adhered part
can be given explicitly. In the following, we non-dimensionalize energies by the bending rigidity. 
Then the total mechanical energy of the adhered part,
$E_{\mathrm{ad}}^{\mathrm{tot}}
=E_{\mathrm{ad}}^{\mathrm{\kappa}}+E_{\mathrm{ad}}^{\mathrm{\sigma}}$,
reads
\begin{equation}
 \frac{E_\mathrm{ad}^\mathrm{tot}}{\kappa}
 =  4\pi (1-\cos\theta) + \pi\frac{R^2}{\lambda^2} (1-\cos\theta)^2\,.
\end{equation}

For the energy of the free parts,
$E_{\mathrm{free}}^{\mathrm{tot}}
=E_{\mathrm{free}}^{\mathrm{\kappa}}+E_{\mathrm{free}}^{\mathrm{\sigma}}$,
one has \cite{foret2014,foret2018mechanosensitivity}
\begin{equation}\label{eq:E_tot_free}
 \frac{E_\mathrm{free}^\mathrm{tot}}{\kappa}
 = \pi \int_0^\infty\left( \dot{\psi} +\frac{\sin \psi }{r} \right)^2 r \, \mathrm{d} s 
 + \frac{ 2 \pi}{\lambda^2} \int_0^\infty (1-\cos \theta) \, r \, \mathrm{d} s\,.
\end{equation}
This energy has to be minimized 
with respect to the free membrane shape at given uptake angle $\theta$.
Together with the geometrical relations between $\psi$, $r$ and $z$, 
one gets the following shape equations 
\cite{seifert1991shape,julicher1994shape,LipoDoeb,DesernoPRE04,foret2014}:
\begin{align}
\ddot{\psi}r^2 \cos \psi  +  \dot{\psi} r \cos^2 \psi +\frac{1}{2} \dot{\psi}^2 
r^2 \sin \psi  \nonumber \\
 - \frac{1}{2} ( \cos^2 \psi+1) \sin \psi
 -\frac{r^2}{\lambda^2} \sin \psi&=0,  
 \nonumber \\
\dot{r}-\cos \psi&=0, \nonumber \\
\dot{z}+\sin \psi&= 0\,.
\label{eq:shape}
\end{align}
This set of ordinary differential equations has 
to be solved with the boundary conditions
\begin{align}
r(0) =R \sin \theta, \;\; \psi(0)&= \theta,\;\;  \psi(\infty) =0, 
\;\; \dot{\psi}(\infty)=0 
\label{eq:boundaryconditions}
\end{align}
and an additional one, $z(\infty)=0$,
where other choices are also possible.

\begin{figure}[t]
    \centering
    \includegraphics[ width=0.48 \textwidth]{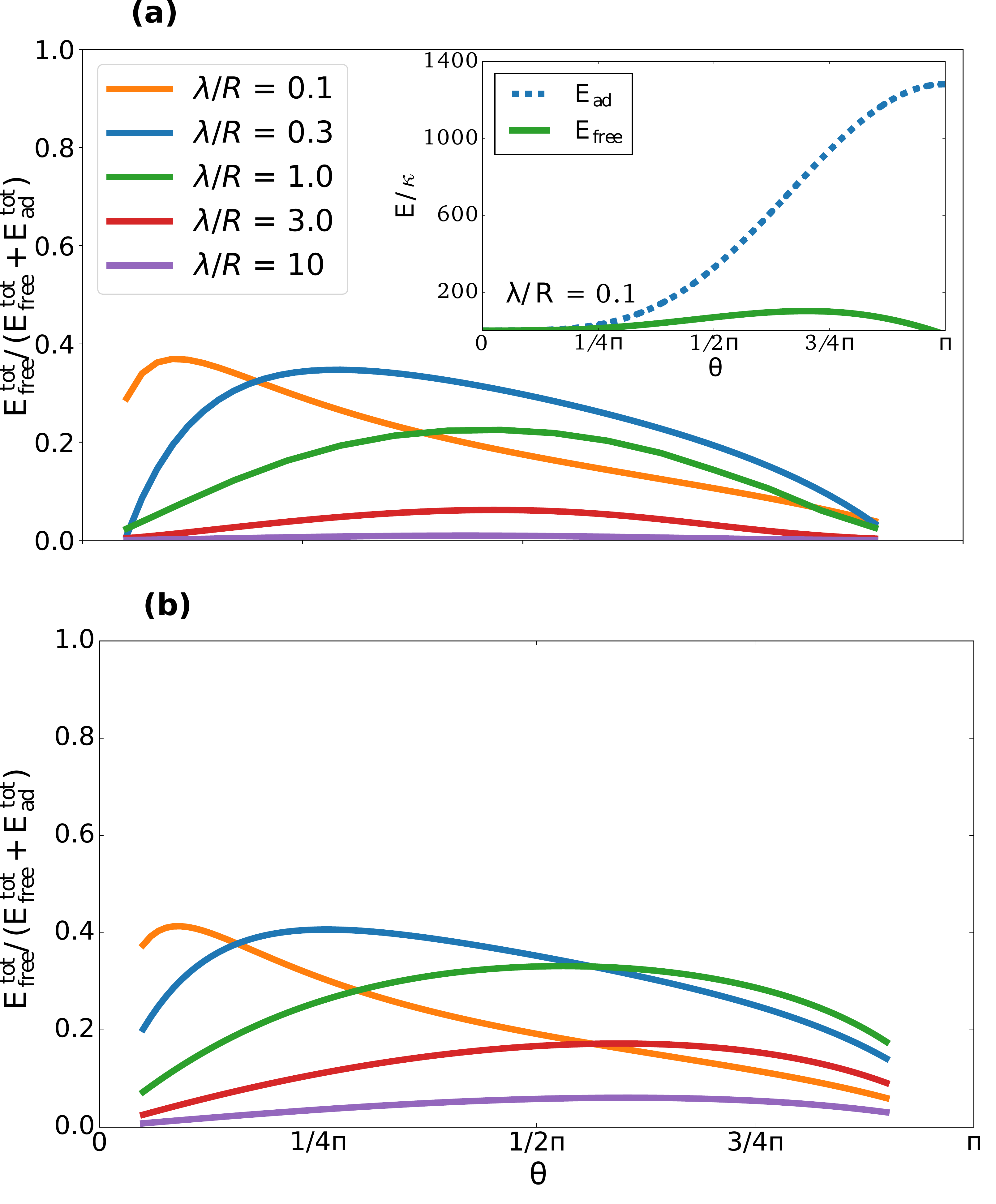}
    \caption{
    (a) The bending and tension energy of the free membrane relative to the total 
    bending and tension energy, i.e.~of the
    adhered and free parts of the membrane for different values of $\lambda/R$.
    To calculate the energy of the free parts we use Eq.~(\ref{eq:Foret_tense}) in the case
    of $\lambda/R<1$ and Eq.~(\ref{eq:Foret_loose}) in the case of  
    $\lambda/R>1$ and the numerical solution for $\lambda = R$.
    The inset shows the bending and tension energy of the adhered and free parts
    of the membrane for $\lambda/R=0.1$.
   (b) The same as in (a) 
        but for the phenomenological description of the free membrane
	given in Eq.~(\ref{free_mb_approx}).
	}
    \label{fig:Figure3}
\end{figure} 

We solved the boundary-value problem for the given system of ODEs
from Eqs.~(\ref{eq:shape}) using a 4-th order collocation algorithm
with matched asymptotics \cite{Solver}. Details can be found in appendix 
\ref{appendix:details_shapeeq}. 
We then evaluated the energy contributions from the free membrane.
In addition, we compared to asymptotic expressions 
that have been given previously \cite{foret2014} 
for the limit of a tense ($ \lambda/R \ll 1$)
and loose membrane ($\lambda / R \gg 1$),
which are also given in appendix \ref{appendix:details_shapeeq}. 
In general, in both limits we get excellent agreement between 
these analytical and our numerical results.

\begin{figure*}[hbt]
    \centering
    \includegraphics[ width=1. \textwidth]{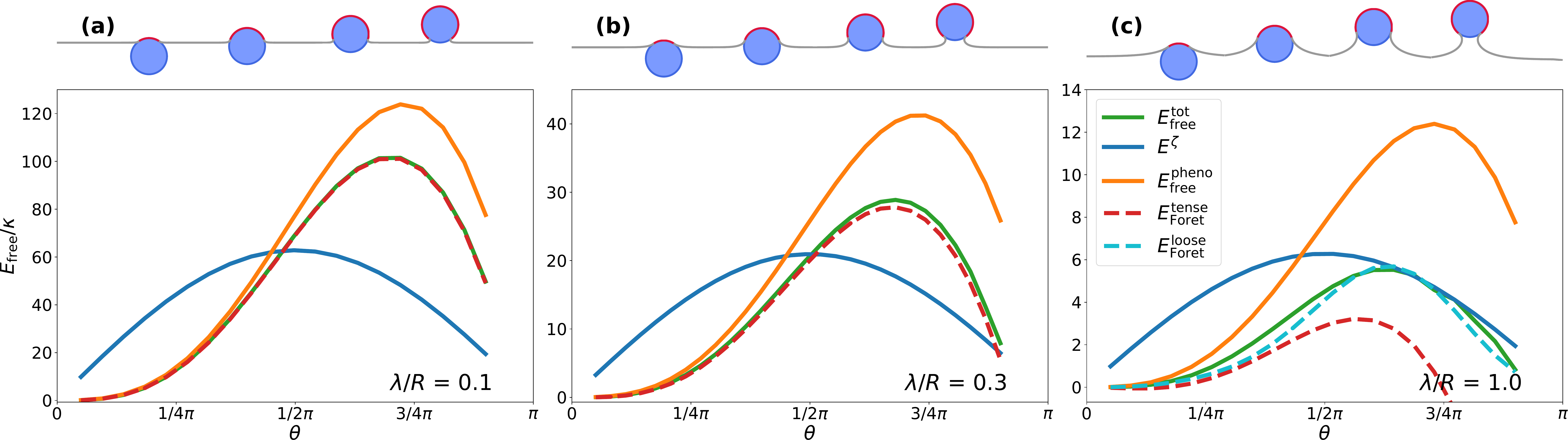}
    \caption{
    Energies of the free membrane for (a) $\lambda/R = 0.1$, (b) $\lambda/R = 0.3$ and (c) $\lambda/R = 1.0 $.
    The numerically calculated total energy of the free membrane is
    shown in green, the line tension approximation $E^\zeta$ is shown in blue, 
    the phenomenological approximation 
    $E_\mathrm{free}^{\mathrm{pheno}}$
    from Eq.~(\ref{free_mb_approx}) is shown in orange 
    and the analytical limits for tense (from Eq.~\ref{eq:Foret_tense}
    \cite{foret2014}) and loose (from Eq.~\ref{eq:Foret_loose}
    \cite{foret2014}) membranes are shown as dashed curves in red and 
    cyan, respectively. (Top) Numerically calculated shapes of the membrane.
    }
    \label{fig:Figure4}
\end{figure*} 

Fig.~\ref{fig:Figure3}(a) shows the mechanical energy of the free membrane
relative to the total mechanical energy for different values of $\lambda/R$.
Here the energies of the free parts were calculated using 
Eq.~(\ref{eq:Foret_tense}) for the tense regime ($\lambda/R<1$),
Eq.~(\ref{eq:Foret_loose}) for the loose regime ($\lambda/R>1$)
and numerically for the intermediate regime ($\lambda = R$).
The analysis demonstrates that in the limit of a
loose membrane  ($\lambda / R \gg 1$) the relative energies of
its free parts are very small.
The underlying reason is that in this case, the membrane 
assumes the shape of a minimal surface \cite{foret2014}
and both bending and tension contributions
become very small. In case of a tense membrane  ($\lambda / R \ll 1$), 
the relative energy of the free membrane plays a role mainly for small angles,
i.e.~for the early uptake dynamics, and then becomes small.
The underlying reason is that in this case, the free membrane is
essentially flat and the neck region is very small compared with the
particle size. Thus the absolute value of the energy of the adhered
part dominates, as demonstrated in the inset of Fig.~\ref{fig:Figure3}(a).
In the intermediate case ($\lambda / R \approx 1$), 
the free membrane contributes mainly at half-uptake, with around $20\%$.
Overall, we conclude that the free membrane contributes 
considerably in the range
\begin{equation}
\lambda / R \sim 0.1-1\, ,
\end{equation}
which is the relevant regime for biological systems \cite{DesernoPRE04}.

The procedure of solving the shape equations is numerically involved 
and makes it difficult to proceed with a comprehensive study of uptake dynamics.
A simple phenomenological expression that
represented the energy of the free membrane well would allow for analytical insight and
for a transparent discussion of the effect of the free membrane on the uptake process.
It has been suggested earlier \cite{LipoDoeb} that the effects of the free
membrane may be seen as an effective line tension.
Considering a spherical particle, this corresponds
to an additional energy contribution 
\begin{equation}\label{line_tension}
 E^\mathrm{\zeta}=  \zeta \mathcal{E} = \zeta 2\pi R \sin \theta\,,
\end{equation}
with $\zeta$ the effective line tension and $\mathcal{E}$
the length of the edge.
However, this simple form has been shown to be strictly true 
only in the double limit of high tension and large uptake angle, where
$E_{\mathrm{free}}^{\mathrm{tot}}$ scales like
a line tension with 
\begin{equation}\label{eq:scaling}
\zeta = \sqrt{\kappa \sigma} \, ,
\end{equation}
as proposed in the works of Deserno \cite{DesernoPRE04}
and Foret \cite{foret2014}.

In order to arrive at such a phenomenological approach, 
we first note that the analytical expressions for the
tense and loose regimes in leading order are
$E_{\mathrm{free}}^{\mathrm{tot}} = \pi \zeta R \theta^2 \sin \theta$
and $E_{\mathrm{free}}^{\mathrm{tot}}  = \pi \zeta R (R/\lambda )\sin^4 \theta$, respectively.
Compared to the expression for a line tension from Eq.~(\ref{line_tension}),
which has a symmetric barrier at $\theta = \pi/2$, these expressions shift the barrier
to higher values of $\theta$. We therefore explored the effects
of analytically simple terms scaling as $\sim \theta \sin \theta$ and $\sim \theta^2 \sin \theta$
and found that the second choice works very well. 
Fig.~\ref{fig:Figure3}(b) shows the results for the relative contribution
of the free membrane for different values of $\lambda / R$ 
with the phenomenological ansatz
\begin{equation}\label{free_mb_approx}
E_\mathrm{free}^{\mathrm{pheno}}= \zeta \pi R \theta ^2 \sin \theta \, .
\end{equation}
They are surprisingly similar to the exact results shown in Fig.~\ref{fig:Figure3}(a).

To further explore the validity of this ansatz, 
in Fig.~\ref{fig:Figure4} we compare the numerically computed total energy
of the free membrane $E_{\mathrm{free}}^{\mathrm{tot}}$ (green) 
to a line tension $E^\mathrm{\zeta}$ (blue) 
and the phenomenological approximation 
$E_\mathrm{free}^{\mathrm{pheno}}$ (orange).
Both for the line tension, defined in Eq.~(\ref{line_tension}),
and the phenomenological approximation, defined in Eq.~(\ref{free_mb_approx}), 
we use the value of $\zeta$ defined in Eq.~(\ref{eq:scaling}).
In the tensed case (Fig.~\ref{fig:Figure4}(a)), 
one sees the excellent agreement between the numerical calculations (green curve)
and the analytical result given as Eq.~(\ref{eq:Foret_tense}) (dashed red). 
Next we note that Eq.~(\ref{eq:Foret_tense}) (dashed red) is a rather
poor description for $\lambda/R=1$ (Fig.~\ref{fig:Figure4}(c)) as it gets negative for
large uptake angles, whereas Eq.~(\ref{eq:Foret_loose}) (dashed cyan) 
is completely off for $\lambda/R<1$ (not shown).
Therefore neither Eq.~(\ref{eq:Foret_tense}) nor Eq.~(\ref{eq:Foret_loose})
is a good description in the relevant domain $\lambda / R \sim 0.1-1$.
As it is also hard to interpolate between the two analytical limits,
the phenomenological approximation is much more convenient.
Very importantly, it performs better than the line tension (cf.~blue curve),
which is not only off quantitatively, but also places the barrier at too small 
values of $\theta$. We conclude that the 
phenomenological approximation $E_\mathrm{free}^{\mathrm{pheno}}$
represents a good description of the qualitative behavior in the biologically
relevant regime and that it works well over the whole range of angles,
different from the ansatz of an effective line tension.

\begin{figure}[b]
    \centering
    \includegraphics[ width=.48 \textwidth]{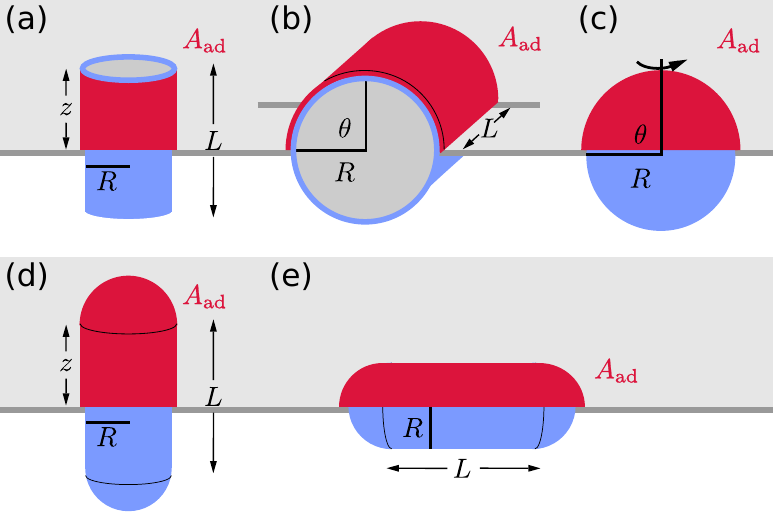}
    \caption{Shapes considered here: cylinders in (a) normal orientation (rocket mode) and 
    (b) parallel orientation (submarine mode), (c) spherical particles,  
    spherocylinders in (d) normal orientation and (e) parallel orientation.
    In a deterministic model, the particle/virus is homogeneously covered with ligands (blue) 
    that can adhere to the cell membrane. The adhered area $A_\mathrm{ad}$ is marked in red.}
    \label{fig:Figure5}
\end{figure}  

\section{Deterministic dynamics}
\label{sec:three}
\subsection{Dynamical equations}

We now discuss the deterministic dynamics of adhesion-mediated 
particle uptake, with a focus on the role of particle shape,
which next to size is the particle feature of largest interest
\cite{vacha2011receptor,huang2013role,dasgupta2014,Bahrami,deng2019entry}.
The focus of this section is an analytical treatment of the dynamics
and therefore here we include the effect of the free membrane only on the level of a line tension.
Our reference shape is always the spherical one due to the dominance of icosahedral viruses.
Cylindrical particles may make contact to the cell membrane 
at any orientation. However, a short (long) cylinder would position 
itself normal (parallel) to the membrane to maximize the initial 
adhered contact area.
It is therefore reasonable to compare these two configurations (orientation normal and parallel) 
to the spherical (icosahedral) shape.
We note that for simplicity here we neglect the top and bottom faces of the cylindrical shape 
and the bending energy of the kinked edges.
In order to study the effect of the edges, we also consider spherocylindrical
particles in normal and parallel orientation (Fig.~\ref{fig:Figure5}).
While our approach is well suited for all these shapes which obey axial 
symmetry, more complex
shapes as for example cubes have to be investigated numerically
\cite{dasgupta2014}.

During uptake the particle adheres to the membrane along the
adhesive area $A_{\mathrm{ad}}$. We describe the progress of uptake 
by the uptake height $z$ for the normal cylinder, 
see~Fig.~\ref{fig:Figure5}(a), or by the uptake angle 
$\theta$ for the parallel cylinder or sphere, see~Fig.~\ref{fig:Figure5}(b), (c).
One can calculate the thermodynamic uptake force by taking the
variation of the energy $E$ with respect to the uptake variable
\begin{equation}
F_\mathrm{up}=-\partial E / \partial x, 
\end{equation}
where $x=z$ or $x=\theta R$.
The uptake force is balanced by a friction force, $F_{\mathrm{up}}=F_{\mathrm{friction}}$.
Here we assume that the dynamics of the uptake process is
dominated by one local timescale. As a lower limit to all possible
choices, we identify the microviscosity $\eta$ of the membrane, 
which is known to dominate uptake times for vesicles
and which has a typical value $\eta=1\ {\rm Pa \; s}$ \cite{agudo2015}. Hence
\begin{equation}
F_\mathrm{friction}=\eta \mathcal{E}(x) \dot{x}. 
\end{equation}
Particle shape will enter our results through the variable edge length $\mathcal{E}(x)$.
If one was interested in another rate-limiting dissipative process for uptake that was also
dominated by one single local time scale acting at the interface between the
adhered and free membrane parts, one could simply
rescale all our results to the desired scale. This however would
not change the relative sequence of uptake and the phase diagrams
presented below. Solving the force balance equation for $\dot{x}$, 
we obtain the dynamic equation for particle uptake. 
In the next paragraphs we briefly summarise the results for the two cylinder
orientations, the sphere and the two spherocylinder orientations.

\begin{figure*}
\centering
\includegraphics[ width=1. \textwidth]{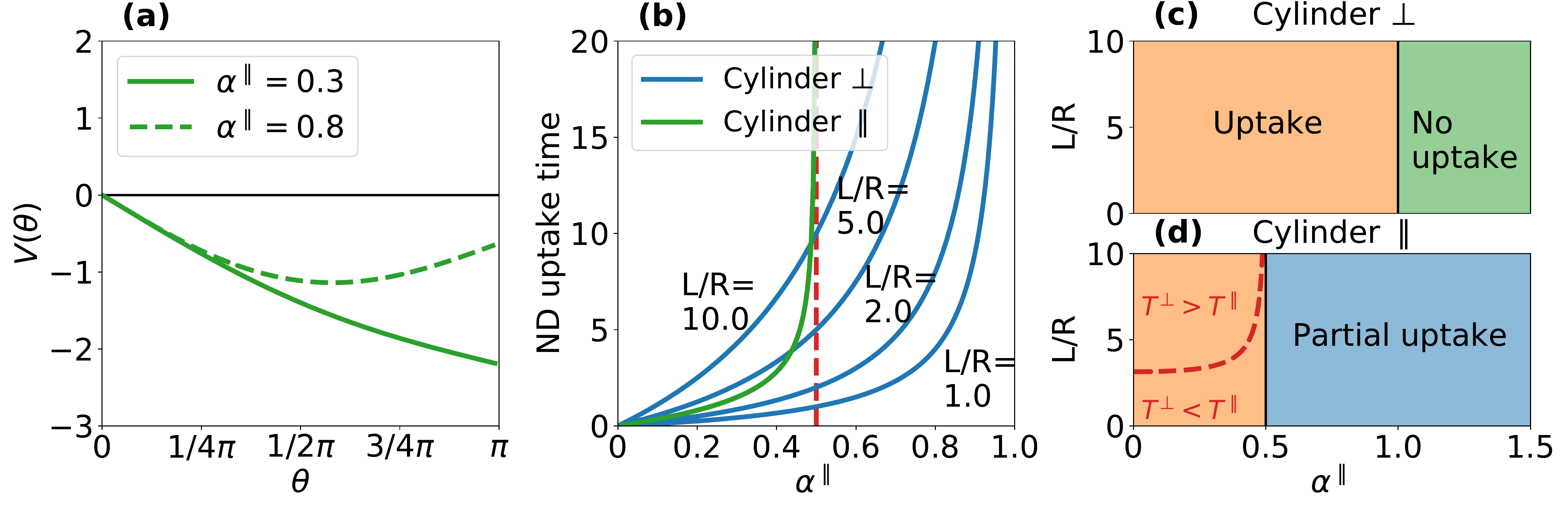}
\caption{\label{fig:Figure6}
Cylindrical uptake. 
(a) Shape of the uptake potential for parallel orientation of the cylinder.
Full uptake is possible for reduced membrane tension 
$\alpha^\parallel<1/2$ (uptake angle $\theta = \pi$ is a boundary minimum, green solid curve).
For $\alpha^\parallel>1/2$, only partial uptake can take place (minimum for finite uptake angle $\theta<\pi$, green dashed).
(b) Non-dimensionalised uptake times of parallel (green) and normal cylinder
(blue) as a function of $\alpha^\mathrm{\parallel}$ at equal radius but different 
aspect ratios (i.e.~different volume).
For the normal cylinder the uptake time increases with aspect ratio whereas it stays constant
for the parallel cylinder.
(c) Dynamical state diagram of normal cylindrical uptake.
Orange region: full uptake, green region: no uptake. 
(d) Dynamical state diagram of parallel cylindrical uptake.
Orange region: full uptake, blue region: partial uptake.
Above the dashed red line the parallel cylinder and below the dashed red line the normal 
cylinder is taken up fastest.
}
\end{figure*}

\subsection{Cylinder with normal orientation $(\bot)$}

For a cylinder (radius $R$, length $L$) oriented normally to the membrane 
one has adhesive area $A_\mathrm{ad}^{\bot}=\mathcal{E}^{\bot} z $,
edge length $\mathcal{E}^\mathrm{\bot}=2 \pi R$  
and mean curvature $H^{\bot}=1/(2R)$, hence
\begin{equation}
E^{\bot}=-W  2 \pi R z + \kappa \frac{\pi z}{R} + 
\sigma 2 \pi R z  
+ \zeta 2 \pi R.
\end{equation}
We again note that the top and bottom surfaces of the cylinder would not
contribute to the uptake force and are neglected here 
for simplicity. The differential equation for the uptake then simply reads
\begin{equation}
\dot{z}=  -\frac{\partial E^{\bot} / \partial z}{\eta \mathcal{E}^{\bot}}.    
\label{eq:det_dyn_NC} 
\end{equation}
One sees that the line tension does not affect the uptake dynamics
as the length of the edge does not change with $z$. Thus the uptake rate is
a constant, $\dot{z}= \nu_{\mathrm{up}}^{\bot}$, with
\begin{equation}
\nu_{\mathrm{up}}^{\bot}=\nu_{\mathrm{w}}^{\bot}-
\nu_{\mathrm{\kappa}}^{\bot}-\nu_{\mathrm{\sigma}}^{\bot}
=W/ \eta -\kappa /(2  R^2 \eta)-\sigma/ \eta. 
\end{equation}

In case $\nu_{\mathrm{w}}^{\bot}$ overcomes the counteracting terms
$\nu_{\mathrm{\kappa}}^{\bot}+\nu_{\mathrm{\sigma}}^{\bot}$ from
bending and tension, uptake progresses at constant speed
and the uptake time is given by $T_{\mathrm{det}}^{\bot}
=L/ \nu_{\mathrm{up}}^{\bot}$.
Otherwise, the particle does not get taken up at all ($\nu_{\mathrm{up}}^{\bot}\le0$).
Partial uptake will never occur.
The critical radius (at which the uptake time diverges) is given by
$
R_{\mathrm{crit}}^{\bot}=\sqrt{\kappa/(2(W- \sigma))}.
$
The optimal radius (for which uptake is fastest) is given by
$
R_{\mathrm{*}}^{\bot}=
\sqrt{3}R_{\mathrm{crit}}^{\bot}.
$
 
Below we will compare all particle shapes 
at equal volume and equal radius because it is a natural question
to ask which shape performs best at a given volume of
transported cargo.
Taking the sphere as the reference shape, 
the normally oriented cylinder then has a length $L=4 R / 3$. 

\subsection{Cylinder with parallel orientation $(\parallel)$}

Completely analogously, one obtains the energy for the parallel cylinder, now as 
a function of the uptake angle (see Fig.~\ref{fig:Figure5}(b)) 
\begin{align}
E^{\parallel}=-W \theta 2 L R  + \kappa \frac{\theta  L}{R} 
+ \sigma(\theta-\sin \theta ) 2LR  
+ \zeta 2 L \, .
\end{align}
Again the bottom and top faces of the cylinder are neglected for simplicity.
Then, the dynamic equation for uptake is given by
\begin{equation}
\dot{\theta}=\nu_{\mathrm{up}}^{\parallel}-
\nu_{\mathrm{\sigma}}^{\parallel} (1-\cos \theta ),
\label{eq:det_dyn_PC}
\end{equation}
where $\nu_{\mathrm{up}}^{\parallel}=\nu_{\mathrm{w}}^{\parallel}
-\nu_{\mathrm{\kappa}}^{\parallel}=W/ (R \eta)
-\kappa / (2 R^3 \eta)$ and $\nu_{\mathrm{\sigma}}^{\parallel}
=\sigma/(R \eta)$.
Again, the line tension does not affect the uptake dynamics.

It is insightful to non-dimensionalize the dynamic equation by introducing the characteristic 
time $1/\nu_{\mathrm{up}}^{\parallel}$ to get 
\begin{equation}\label{eqcylrescaled}
\frac{\mathrm{d} \theta}{\mathrm{d} \tau} = 1 - \alpha^{\parallel} (1-\cos \theta ).
\end{equation}
We will assume $\nu_{\mathrm{w}}^{\parallel}
>\nu_{\mathrm{\kappa}}^{\parallel}$ (since otherwise there is no uptake anyways) and hence
the reduced membrane tension
\begin{equation}
 \alpha^\parallel=\nu_\sigma^\parallel / \nu_{\mathrm{up}}^{\parallel}
= \frac{2 \sigma R^2} {2 W R^2-\kappa}
\end{equation}
is positive and uptake is expected to be the faster the smaller $\alpha^\parallel$ is. 
Eq.~(\ref{eqcylrescaled}) can be integrated analytically with initial condition
$\theta(t=0)=0$ to obtain $\theta(t)$. 
It is, however, simpler, to write it in potential form,
\begin{equation}
 \frac{\mathrm{d} \theta}{\mathrm{d} \tau} =-\frac{dV(\theta)}{d\theta}\,\,,\,\,\,
 V(\theta)=-\theta+\alpha^\parallel(\theta-\sin\theta)\,,
\end{equation}
highlighting the dynamic behaviour: first,
the slope for small angles is always negative. 
For $0<\alpha^{\parallel} \leq 1/2$ one has a 
boundary minimum at $\theta=\pi$, hence complete uptake
(although the uptake time diverges at $\alpha^{\parallel}=1/2$).
For $\alpha^{\parallel}>1/2$ the minimum is for $\theta<\pi$ and hence one
has only partial uptake, cf.~Fig.~\ref{fig:Figure6}(a). 
Integrating Eq.~(\ref{eqcylrescaled}) leads to the
uptake time 
\begin{equation}
T_{\mathrm{det}}^{\parallel}\approx\frac{\pi}{\nu_{\mathrm{up}}^{\parallel}
\sqrt{1-2 \alpha^{\parallel}}} \, ,
\end{equation}
for $\alpha^{\parallel}<1/2$ 
and to the final partial wrapping angle of
\begin{equation}
\theta(t\rightarrow\infty)=2\arctan \left(\frac{1}{\sqrt{2\alpha^{\parallel}-1}} \right) \, ,
\end{equation}
for  $\alpha^{\parallel}>1/2$. 
The critical radius is determined by $\alpha^{\parallel}=1/2$, yielding
$
R_{\mathrm{crit}}^{\parallel}=\sqrt{\kappa/(2(W- 2\sigma))}.
$

\subsection{Comparison of cylinder with normal and parallel orientation}

To compare the uptake times of the normally and parallelly oriented cylinders, 
we can express both in the reduced membrane tension $\alpha^\parallel$ to get
\begin{equation}
\frac{T^\bot_\mathrm{det}}{R\eta/\sigma}= \frac{L}{R} \frac{\alpha^\parallel}{1-\alpha^\parallel}\,\,\,,\,\,\, 
\frac{T^\parallel_\mathrm{det}}{R\eta/\sigma}=\pi \frac{\alpha^\parallel}{\sqrt{1-2\alpha^\parallel}} \, ,
\label{eq:nd_tup_parallel}
\end{equation}
where $R \eta / \sigma$ is again a characteristic time scale.

Fig.~\ref{fig:Figure6}(b) shows the rescaled uptake times for the normal (blue) and 
parallel cylinder (green) as a function of $\alpha^\parallel$
at equal radius but different aspect ratios $L/R$ (i.e.~different volume).
While for the parallel cylinder the uptake time is a constant, 
for the normally oriented cylinder it naturally increases with length and hence the aspect ratio.  
Consequently, the normal cylinder is faster only as long as it is rather short,
explicitly as long as
\begin{equation}
\frac{L}{R}\le \pi  \frac{1 - \alpha^\parallel}{\sqrt{1-2 \alpha^\parallel}}\,.
\end{equation}
The optimal uptake orientation of a cylinder hence depends on the aspect ratio.

Fig.~\ref{fig:Figure6}(c) and (d) show the dynamical state diagrams
for normal and parallel cylinder uptake
as a function of aspect ratio and $\alpha^\parallel$.
While normal cylinders are either taken up completely or not at all, 
parallel cylinders can also be taken up partially.
We note that the state diagram for the the parallel cylinder 
is in agreement with the one which was previously computed by 
Mkrtchyan and coworkers \cite{mkrtchyan2010adhesion}.
In Fig.~\ref{fig:Figure6}(d), we also investigate the speed of uptake.
Below the dashed red line the normal cylinder is taken up fastest, whereas
above the dashed red line the parallel cylinder is taken up fastest.

\subsection{Sphere $(\circ)$}

For a sphere the total energy reads
\begin{align}
 E^{\circ}= &\left (- W 2 \pi R^2 + \kappa 4 \pi +\sigma \pi R^2 
 (1-\cos \theta)  \right ) \nonumber\\
 & \times (1-\cos \theta )+\zeta 2 \pi R \sin  \theta \, ,
 \label{eq:energy_sphere}
\end{align}
and hence the differential equation for uptake reads
\begin{align}
\dot{\theta}^{\circ} 
&= \nu_{\mathrm{up}}^{\circ}-\nu_{\sigma} ^{\circ}(1-\cos \theta ) -
 \nu_{\zeta}^{\circ} \cot{\theta} \, ,
\label{eq:det_ODE}
\end{align}
where we have introduced  
$\nu_\mathrm{up}^{\circ}=  \nu_{\mathrm{w}}^{\circ}
-\nu_{\mathrm{\kappa}}^{\circ}= W/(R \eta)-2 \kappa/(R^3 \eta)$, 
$\nu_{\sigma}^{\circ}=\sigma/(R\eta)$ and
$\nu_{\zeta}^{\circ}=\zeta/(R^2 \eta)$.

\begin{figure}[t]
\centering
\includegraphics[ width=.48\textwidth]{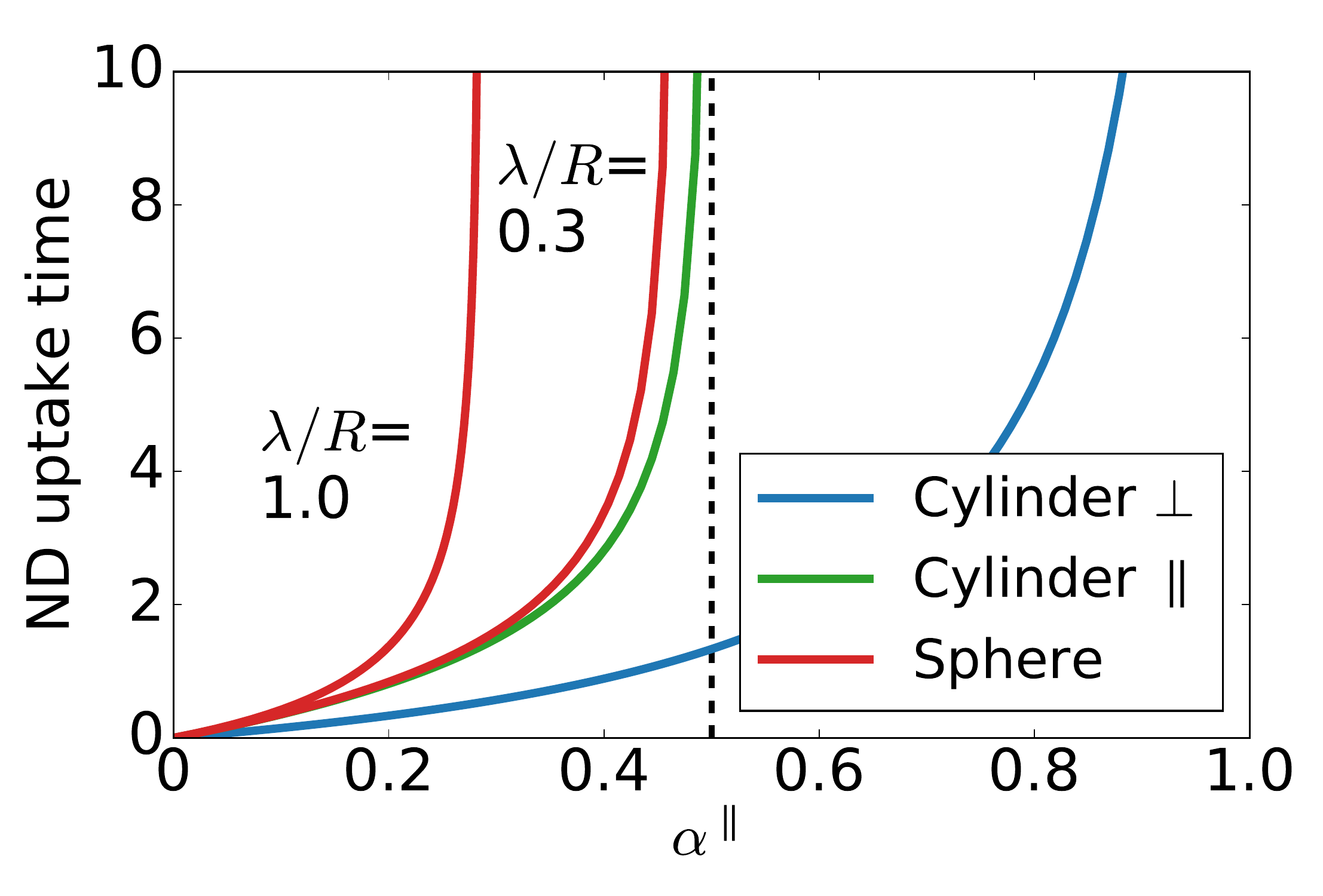}
\caption{
\label{fig:Figure7}
Non-dimensional uptake times for normal cylinder (blue), parallel cylinder (green)
and sphere (red) for $\lambda/R=0.3$ and
$\lambda/R=1.0$ as a function of $\alpha^\parallel$
and for equal volume at equal radius.
}
\end{figure}

Neglecting line tension, the dynamic equation has the same form as for the
parallel cylinder, albeit with different expressions for the rates.
Introducing the reduced membrane tension
\begin{equation}
 \alpha^\circ=\nu_{\mathrm{\sigma}}^\circ/\nu_{\mathrm{up}}^\circ=
\frac{\sigma R^2} {WR^2-2\kappa}\,,
\end{equation}
one hence again has for the uptake time
$T_{\mathrm{det}}^{\circ} \approx \pi/(\nu_{\mathrm{up}}^{\circ}
\sqrt{1-2 \alpha^{\circ}})$, 
implying a critical radius 
$R_{\mathrm{crit}}^\circ=\sqrt{2\kappa / (W- 2\sigma)}$. 

To compare to the uptake times for the normal and parallel cylinder 
at equal volume and radius, we can again express the uptake time as a function of 
$\alpha^{\parallel}$ by means of
\begin{equation}
 \alpha^\circ=\frac{1}{\frac{1}{\alpha^\parallel}-\frac{3 \lambda^2}{2 R^2}} \, .
\label{eq:nd_tup_sphere}
\end{equation}
A comparison of the uptake times of the two cylinder orientations and the sphere 
is shown in Fig.~\ref{fig:Figure7}.
In case of the particle being large compared to the characteristic length scale of the membrane
($\lambda/R\leq1$, tense membrane case), the 
parallel cylinder and the sphere have very similar  uptake times, 
as also evident from Eq.~(\ref{eq:nd_tup_sphere}). For instance,
for $\lambda/R\le0.1$, the uptake time for the sphere
almost coincides with the green curve in Fig.~\ref{fig:Figure7}.
For smaller particles or a looser membrane, the uptake times increasingly separate, 
and the sphere is increasingly disfavored, as seen by the red curves in Fig.~\ref{fig:Figure7}.
In general we thus find that spheres are taken up slower than cylinders
in the deterministic description \cite{frey2019}.

\subsection{Spherocylinder}

We now consider a spherocylindrical particle either in normal or parallel orientation
with respect to the membrane. For the normal orientation the uptake time is
given by the sum of the uptake times for a sphere and for a normal
cylindrical particle

\begin{equation}
T_\mathrm{det}^{\,\cap }=T_\mathrm{det}^{\circ } + T_\mathrm{det}^{\bot} \, .
\end{equation}

We note that at equal volume the radius of a spherocylinder ($R^{\cap}$)
is always smaller compared to the radius of a sphere ($R^{\circ}$). 
Let us first assume that we compare a sphere and spherocylinder, which 
both have smaller radii compared to the optimal radius of the sphere at equal 
volume. In this case, it holds that 
$T_\mathrm{det}^{\,\cap }(R^{\cap})=T_\mathrm{det}^{\circ }(R^{\cap}) +
T_\mathrm{det}^{\bot} (R^{\cap})
>T_\mathrm{det}^{\circ }(R^{\circ})$, since 
$T_\mathrm{det}^{\bot}(R^{\cap}) > T_\mathrm{det}^{\circ }(R^{\circ })-
T_\mathrm{det}^{\circ}(R^{\cap})< 0$ (i.e.~the uptake time of a sphere
decreases below the optimal radius with increasing size).

\begin{figure}[t]
\centering
\includegraphics[ width=0.48 \textwidth]{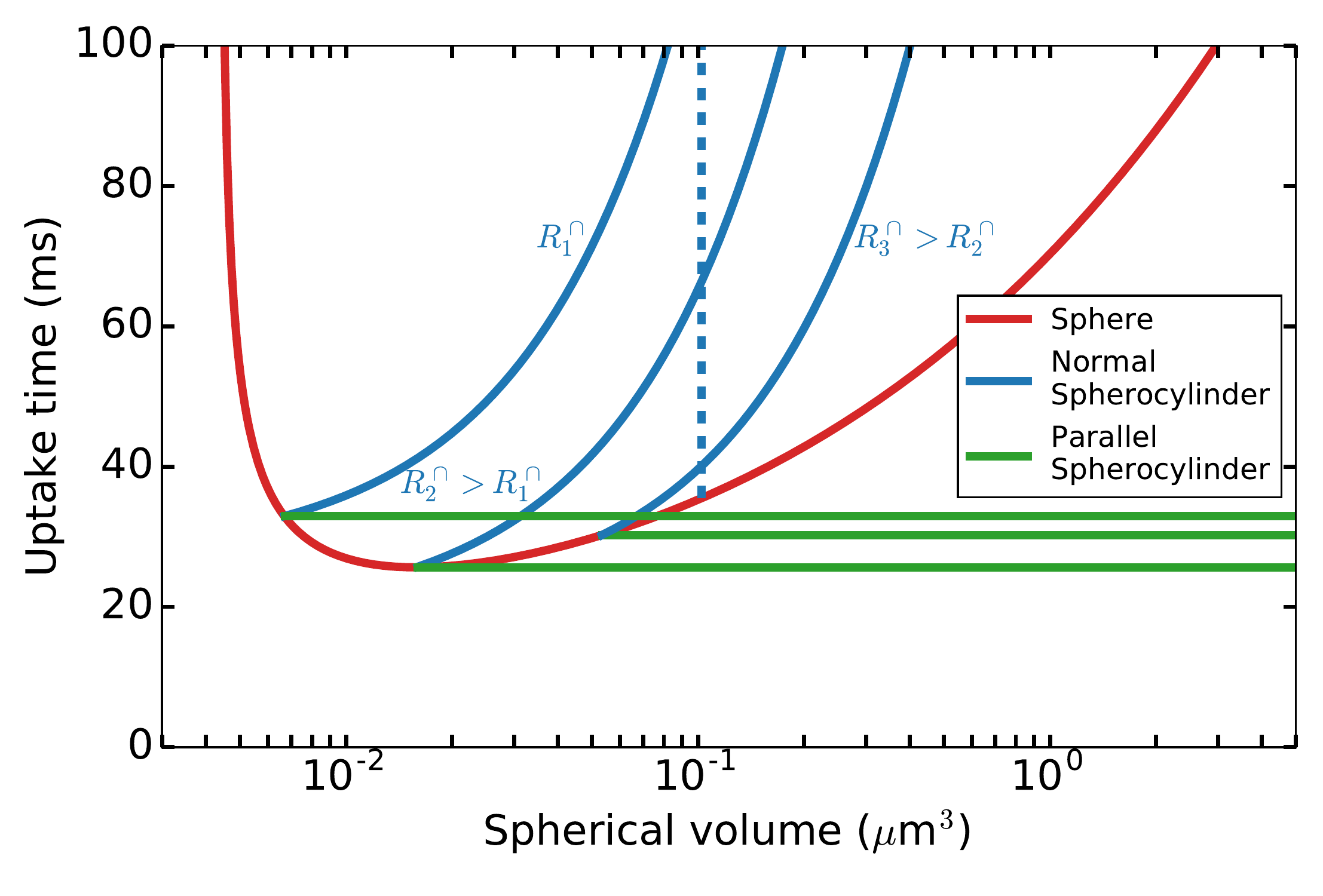}
\caption{
Uptake time as a function of spherical volume for the sphere (solid red) and
spherocylinders with different radii in normal (solid blue) and parallel (solid green) 
orientation. The dashed blue line marks the uptake times of normal spherocylindrical
particles at equal volume. 
Starting from the red curve, the length of the cylindrical part increases while
reducing the particle radius.
}
\label{fig:Figure8}
\end{figure}

To get insight into the general situation, we consider in Fig.~\ref{fig:Figure8} 
the uptake time as a function of volume of three 
normal spherocylinders (solid blue lines) with fixed but different
radii. We compare the spherocylinders to a sphere with equal volume 
(solid red). At vanishing cylindrical length $L$ (cf.~Fig.~\ref{fig:Figure5}(d))
the uptake times of the spherocylinders and sphere are of course identical. 
For increasing volume, i.e.~increasing cylindrical length we find that the sphere 
is always taken up faster compared to the normal spherocylinders. 
In the case when the radius and length are changed at equal volume,
one moves on vertical lines through the diagram (dashed blue line). 
By reducing the cylindrical length, the particle radius has to increase,
while keeping the volume fixed. Upon reducing the length one finally gets back onto the uptake
curve of a sphere (which is of course identical to a spherocylinder 
with vanishing cylindrical length). 
Since this argument can be repeated for all radii, the spherocylinder is
always taken up slower compared to the sphere at equal volume.

\begin{figure}[b]
\centering
\includegraphics[ width=.48 \textwidth]{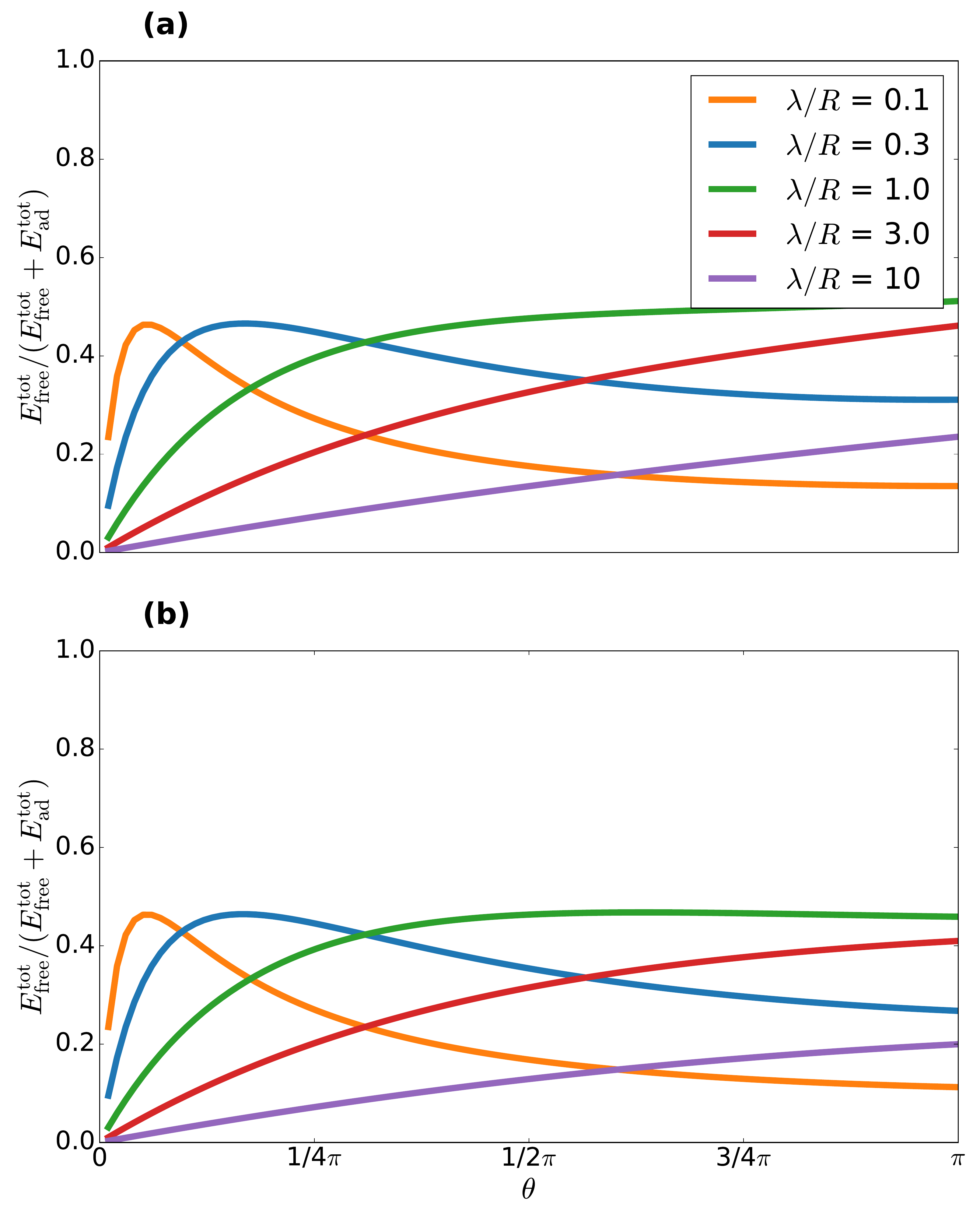}
\caption{
(a) The bending and tension energy of the free membrane 
of a cylindrical particle in parallel orientation relative to the total 
bending and tension energy, i.e.~of the
adhered and free parts of the membrane for different values of $\lambda/R$.
(b) Similar to (a) but now for the phenomenological description of 
the free membrane parts.}
\label{fig:Figure9}
\end{figure}

\begin{figure*}
\centering
\includegraphics[ width=1. \textwidth]{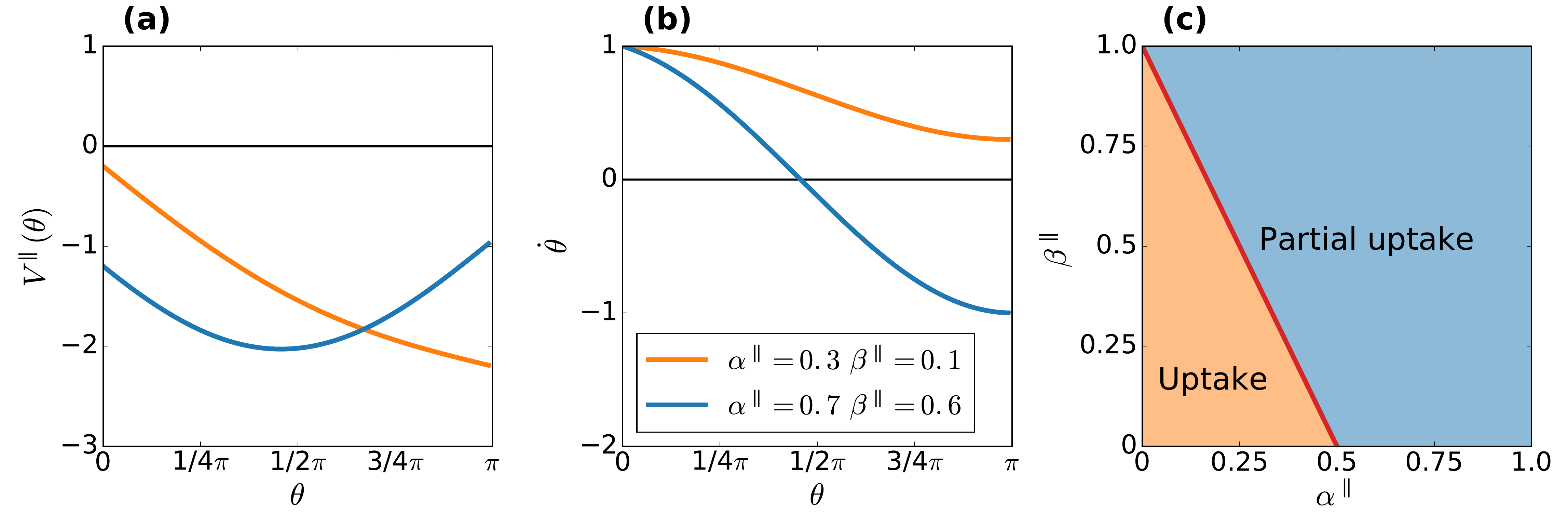}
\caption{
Deterministic uptake dynamics of a cylinder in parallel orientation including
the exact solution for the free membrane.
(a) Uptake potential, leading to uptake (orange) and partial uptake (blue).
Parameters are given in (b).
(b) Phase portrait ($\dot\theta$ vs.~$\theta$) for different parameter
values corresponding to uptake (orange) and partial uptake (blue).
(c) Dynamical state diagram of the final steady states as a function of 
$\beta^\parallel$ vs.~$\alpha^\parallel$. Below the red curve the cylinder is taken up
completely (orange region) and above we find only partial uptake (blue region).}
\label{fig:Figure10}
\end{figure*} 

For the spherocylindrical particle in parallel orientation the uptake time
is given by the slowest mode as the spherical part and the cylindrical part 
are taken up simultaneously. As the uptake time for the parallel cylindrical 
particle is always faster than the uptake time of a spherical particle
at equal volume and radius the uptake time of the 
spherocylindrical particle in parallel orientation is given by the 
uptake time of the spherical particle alone
\begin{equation}
T_\mathrm{det}^{\,\supset }=T_\mathrm{det}^{\circ } \, .
\end{equation}
Hence, the uptake time of the spherocylindrical particle in parallel orientation
is constant while adapting the volume by only increasing the length of
the cylindrical part.
In Fig.~\ref{fig:Figure8} the uptake times for a spherocylindrical 
particles in parallel orientation with fixed but different spherical radii are shown in solid green.
While for a sphere the relation between volume and radius is unambiguous,
for a spherocylinder there exist different combinations of radii and lengths with the same volume.
While the lowest green curve corresponds to a parallel spherocylinder with
optimal spherical radius, the curve starting at smaller (larger) volume corresponds 
to a parallel spherocylinder with smaller (larger) radius. We see that
spherical particles present clear optima in terms of uptake times at intermediate volumes, 
while spherocylinders are faster for large volumes, because there additional volume 
does not increase their uptake times when implemented by increased length.

To conclude, at equal volume a normal spherocylinder
is taken up slower compared to a sphere, independent of length.
For a parallel spherocylinder the order of uptake depends on the
aspect ratio. For the same volume and aspect ratio parallel spherocylinders
are taken up faster compared to normal spherocylinders.
Comparing the uptake times of a (normal or parallel) cylinder 
to a (normal or parallel) spherocylinder at equal volume and radius, 
we find that cylinders are always taken up faster. The reason is that 
it is more time consuming to wrap the spherical caps compared to the 
cylindrical parts with the same volume.

\section{State diagrams for parallel cylinder}

\label{sec:four}
\subsection{Free membrane energies}
We now explicitly consider the effect of the free membrane.
We first note that for a cylindrical particle 
in normal orientation, the shape equations are similar to the case of a 
spherical particle with $\theta=\pi/2$. 
However, the energy of the free membrane is a constant since it does not depend on the 
invagination depth $z$. Therefore, the free membrane will not contribute to the dynamics
of particle uptake. The first non-trivial case therefore is the parallel
cylinder. In the next section, we will then discuss the spherical particle.

For the parallel cylinder the energy of the free membrane 
compared to the flat case can be written as \cite{mkrtchyan2010adhesion}
\begin{equation}
\frac{E_\mathrm{free}^\mathrm{tot \parallel}}{\kappa} 
= \frac{8 L}{\lambda} \left (1-\cos \frac{\theta}{2} \right ) 
\, .
\end{equation}
Details can be found in appendix 
\ref{appendix:free_membrane_parallel_cylinder}. 
Considering the free membrane on both cylinder sides, the scaling of the
free energy is then given by 
$E_\mathrm{free}^\mathrm{tot \parallel}  \sim \zeta L \theta^2=
E_\mathrm{free}^\mathrm{pheno \, \parallel} $, which is
the analogous of the phenomenological approach of 
Eq.~(\ref{free_mb_approx}) for spheres.
The energy of the free membrane relative to the energy of both
the free and adhered membrane parts are shown in 
Fig.~\ref{fig:Figure9}(a) for different values
of $\lambda/R$. In Fig.~\ref{fig:Figure9}(b)
the same is shown for the phenomenological description. 
The agreement is very good and we conclude that
the phenomenological approach works very well in this case.

\subsection{Dynamics with free membrane}

We now study the uptake dynamics for the parallel cylinder
including the free membrane. 
The dynamical equation in non-dimensional form
reads 
\begin{equation}
\frac{\mathrm{d} \theta}{\mathrm{d} \tau} =1 - 
\alpha^\parallel (1-\cos \theta ) - \beta^\parallel \sin{\theta/2} \, ,
\label{eq:det_ODE_cylinder_parallel}
\end{equation}
with $\tau=t \nu_{\mathrm{up}}^{\parallel}$
and the reduced line tension 
\begin{equation}
\beta^\parallel=\frac{\nu_\zeta^\parallel}{ \nu_\mathrm{up}^\parallel} = 
\frac{4 \sqrt{ \kappa \sigma} R}
{2 WR^2-\kappa} 
\;\;\;
\mathrm{and} 
\;\;\; 
\nu_\zeta^\parallel=\frac{2\sqrt{\kappa \sigma}}{\eta R^2} \, ,
\end{equation}
which incorporates the effect of the free membrane. Importantly, the
uptake dynamics does not depend on the length of the parallel
cylinder.

\begin{figure*}
\centering
\includegraphics[ width=1. \textwidth]{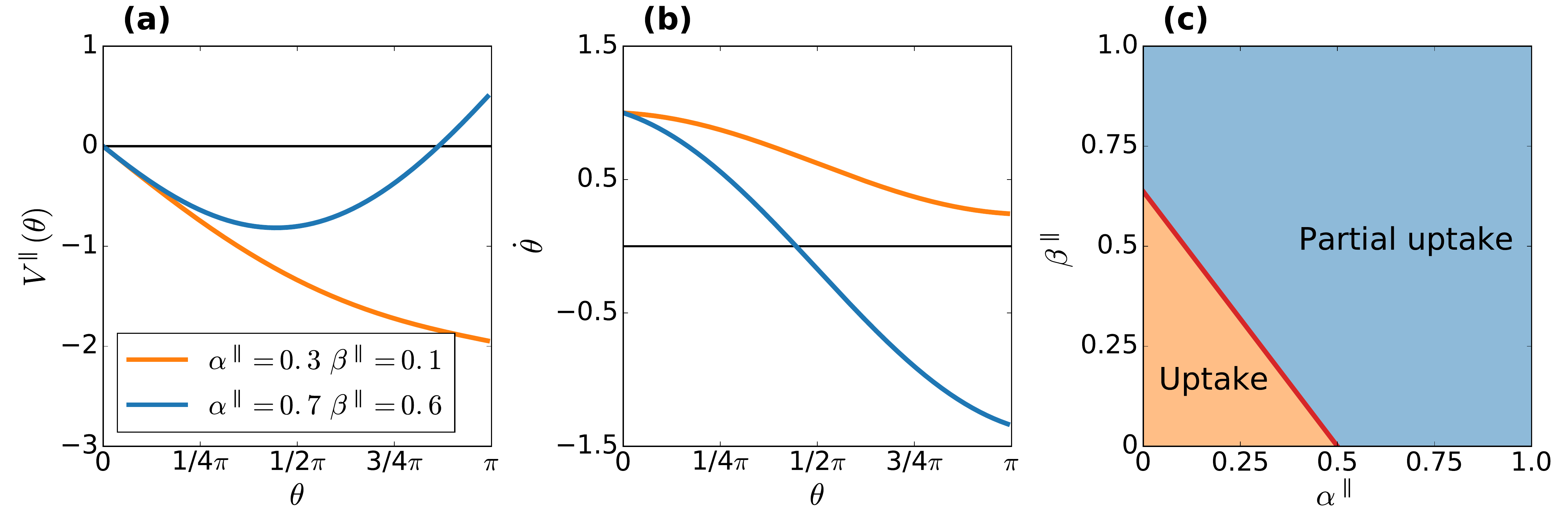}
\caption{
Deterministic uptake dynamics of a cylinder in parallel orientation including
the free membrane by the phenomenological description.
(a) Uptake potential for different parameter values corresponding 
to uptake (orange) and partial uptake (blue).
(b) Phase portrait ($\dot\theta$ vs.~$\theta$). 
Parameters are given in (a).
(c) Dynamical state diagram of the final steady states as a function of 
$\beta^\parallel$ vs.~$\alpha^\parallel$. Below the red curve the cylinder is taken up
completely (orange region) and above we find only partial uptake (blue region).}
\label{fig:Figure11}
\end{figure*}

In potential form, the dynamics 
$\frac{\mathrm{d} \theta}{\mathrm{d} \tau} =-\frac{dV
(\theta)}{d\theta}$ follows from
\begin{equation}
 V(\theta)=-\theta+\alpha^\parallel(\theta-\sin\theta)
 -2\beta^\parallel \cos \frac{\theta}{2}\,.
\end{equation}
This potential is shown in Fig.~\ref{fig:Figure10}(a).
One either has a boundary minimum corresponding to full uptake (orange) 
or a minimum in between corresponding to partial uptake (blue), depending
on the parameter choice.
From the shape of the potential and the phase portrait in 
Fig.~\ref{fig:Figure10}(b) we can deduce
the dynamical state diagram of uptake in the $\beta^\parallel$-$\alpha^\parallel$-plane in
Fig.~\ref{fig:Figure10}(c)
for a parallel cylinder.
We find that the parallel cylinder either gets taken up completely (orange) or 
only partially (blue).
The boundary between these two states is given when 
Eq.~(\ref{eq:det_ODE_cylinder_parallel}) becomes zero. As the second
and third term in Eq.~(\ref{eq:det_ODE_cylinder_parallel}) become minimal
for $\theta=\pi$, we find the boundary between the partial and full uptake 
state when 
\begin{equation}
\frac{\mathrm{d} \theta}{\mathrm{d} \tau}
\bigg |_{\theta^\parallel=\pi}=0
\rightarrow \beta^\parallel=1-2 \alpha^\parallel \, .
\end{equation}

\subsection{Dynamics with phenomenological approach}

We now consider the phenomenological ansatz adapted to the case of a parallel cylinder.
In general, the energy of the phenomenological approach can be written as 
$E_\mathrm{free}^\mathrm{pheno}=1/2 \zeta \theta^2 \mathcal{E}$, 
where $\mathcal{E}$ is the length of the line that connects the adhered membrane 
and the free membrane. For the parallel cylinder we have $\mathcal{E}^\parallel=2L$
and thus $E_\mathrm{free}^\mathrm{pheno \, \parallel}=\zeta L \theta^2$,
where $\zeta=\sqrt{\kappa \sigma}$.
The dynamics for the phenomenological approach is then given by
\begin{equation}
\frac{\mathrm{d} \theta_\mathrm{pheno}}{\mathrm{d} \tau} =1 - 
\alpha^\parallel (1-\cos \theta ) - \frac{\beta^\parallel}{2} \theta \, .
\label{eq:det_ODE_cylinder_parallel_pheno}
\end{equation}
In potential form, the dynamics now reads
\begin{equation}
 V_\mathrm{pheno}(\theta)=-\theta+\alpha^\parallel(\theta-\sin\theta)
 +\frac{\beta^\parallel}{4} \theta^2\,.
\end{equation}
The boundary between these two states is given when 
Eq.~(\ref{eq:det_ODE_cylinder_parallel_pheno}) becomes zero. As the second term in Eq.~(\ref{eq:det_ODE_cylinder_parallel_pheno}) becomes minimal
for $\theta=\pi$ we find
\begin{equation}
\frac{\mathrm{d} \theta_\mathrm{pheno}}{\mathrm{d} \tau}
\bigg |_{\theta^\parallel=\pi}=0
\rightarrow \beta^\parallel=\frac{2-4 \alpha^\parallel}{\pi}\ .
\end{equation}
The potential in Fig.~\ref{fig:Figure11}(a), the uptake dynamics 
in Fig.~\ref{fig:Figure11}(b) and the state diagram of uptake
in Fig.~\ref{fig:Figure11}(c) for the dynamics in the phenomenological
description can be easily compared to the exact case of the free membrane in
Fig.~\ref{fig:Figure10}. We see that the dynamics are very similar.

\begin{figure*}
\centering
\includegraphics[ width=1. \textwidth]{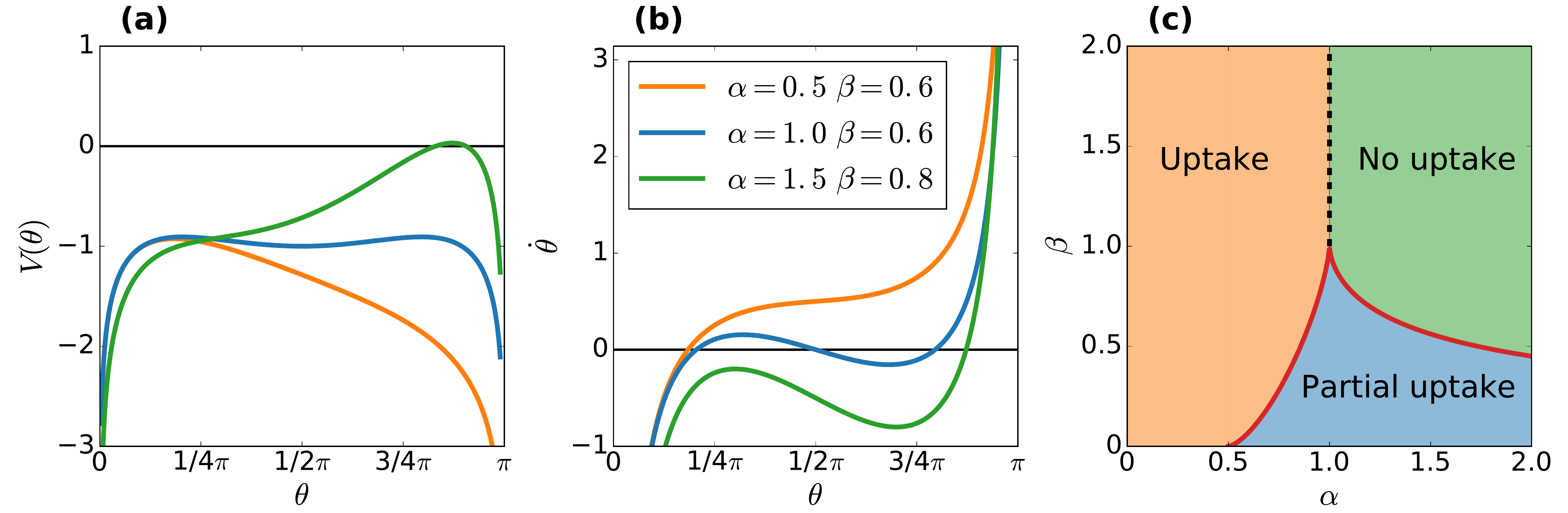}
\caption{
Deterministic uptake dynamics of spherical particles with line tension.
(a) Uptake potential, leading to uptake (orange), partial uptake (blue) and no uptake (green).
Parameters are given in (b).
(b) Phase portrait ($\dot\theta$ vs.~$\theta$) for different parameter
values corresponding to uptake (orange), partial uptake (blue) and 
no uptake (green). 
(c) Dynamical state diagram of the final steady states as a function of $\beta$ vs.~$\alpha$.
Below the red curve the sphere is taken up only partially (blue region) 
and above we find either uptake (orange region) or no uptake (green region).
}
\label{fig:Figure12}
\end{figure*}

\section{State diagrams for sphere}
\label{sec:five}

\subsection{Line tension}

We now study the uptake dynamics for the sphere by 
including the free membrane by a line tension or by a
phenomenological ansatz. We first consider a line tension.
The dynamical equation in non-dimensional form reads 
\begin{equation}
\frac{\mathrm{d} \theta}{\mathrm{d} \tau} =1 - 
\alpha (1-\cos \theta ) - \beta \cot{\theta} \, ,
\label{eq:det_ODE_ciricle}
\end{equation}
with $\tau=t \nu_{\mathrm{up}}^{\circ}$ and
the reduced line tension 
\begin{equation}
\beta=\nu_{\zeta}^\circ / \nu_\mathrm{up}^\circ=
\frac{\zeta R} {WR^2-2\kappa} \, ,
\end{equation}
and the reduced surface tension $\alpha=\alpha^\circ$ as before.
In potential form, the dynamics reads
$\frac{\mathrm{d} \theta}{\mathrm{d} \tau} =-\frac{dV(\theta)}{d\theta}$, now with
\begin{equation}
 V(\theta)=-\theta+\alpha(\theta-\sin\theta)+\beta\ln(\sin\theta)\,.
\end{equation}
From the shape of the potential, visualized in Fig.~\ref{fig:Figure12}(a), one can clearly see that the line tension
creates divergences towards $-\infty$ for both $\theta=0$ and $\theta=\pi$. 
The divergence at small $\theta$ is well known from classical nucleation theory
and implies that a fluctuation is required to start the process. 
The divergence at large $\theta$ reflects the fact that a line
tension accelerates the process once the equator is passed.

Since the potential varies between $-\infty$ and $-\infty$ for $0\leq\theta\leq\pi$,
it must have at least one maximum, corresponding to an unstable steady state.
In case this is the only steady state, we now assume that the initial fluctuations bring
the system over the initial barrier and the result will be full uptake.
However, as a function of reduced membrane and line tension 
$\alpha$ and $\beta$, additional extrema in the potential
and hence additional steady states can emerge.
We identify two scenarios. If the 
potential displays two maxima separated by a minimum,
then we have another steady state that we interpret as partial uptake.
If the potential displays a saddle for values of $\theta$ smaller then those for the maximum,
then we conclude that no uptake is possible since $\dot\theta<0$ except for very large $\theta$.
Fig.~\ref{fig:Figure12} visualizes our three scenarios for the dynamics 
both via the potential (a) and using the phase portrait $\dot\theta$ vs.~$\theta$ (b).

The steady states of Eq.~(\ref{eq:det_ODE_ciricle}) 
as a function
of $\alpha$ and $\beta$ can be studied analytically.
In fact, their number  can change only
if the curvature of a given stationary point changes sign.
One can reformulate the problem to find the particular value of $\beta$ where 
the two functions
$f(\theta_\mathrm{ss})=1 - \alpha (1-\cos \theta_\mathrm{ss} )$ and 
$g(\theta_\mathrm{ss})=\beta \cot \theta_\mathrm{ss}$
are equal, $f(\theta_\mathrm{ss})=g(\theta_\mathrm{ss})$ 
(defining a steady state), and where their derivatives are equal,
$ f'(\theta_\mathrm{ss})=g'(\theta_\mathrm{ss})$ 
(defining the change in the sign of the curvature).
The second condition implies that
$\sin^3 \theta_\mathrm{ss} =\beta / \alpha$
and insertion into the first equation results in
\begin{equation}
\beta (\alpha) =\alpha \left [ 1- \left (  \frac{1}{\alpha}-1\right )^{\frac{2}{3}} 
\right ]^{\frac{3}{2}}.
\label{eq:bifurcation}
\end{equation}
Fig.~\ref{fig:Figure12}(c) shows the dynamical state diagram  for the uptake
of a sphere in the $\beta$-$\alpha$-plane.
Eq.~(\ref{eq:bifurcation}) is shown as the red curves. 
Below these curves one has partial uptake (blue region) 
and above we find either uptake (orange region) 
or no uptake (green region), according to the definition discussed above.
Note that the red curve tends to $\alpha=1/2$ for $\beta\rightarrow0$,
as expected.
 
 \begin{figure*}
\centering
\includegraphics[ width=1. \textwidth]{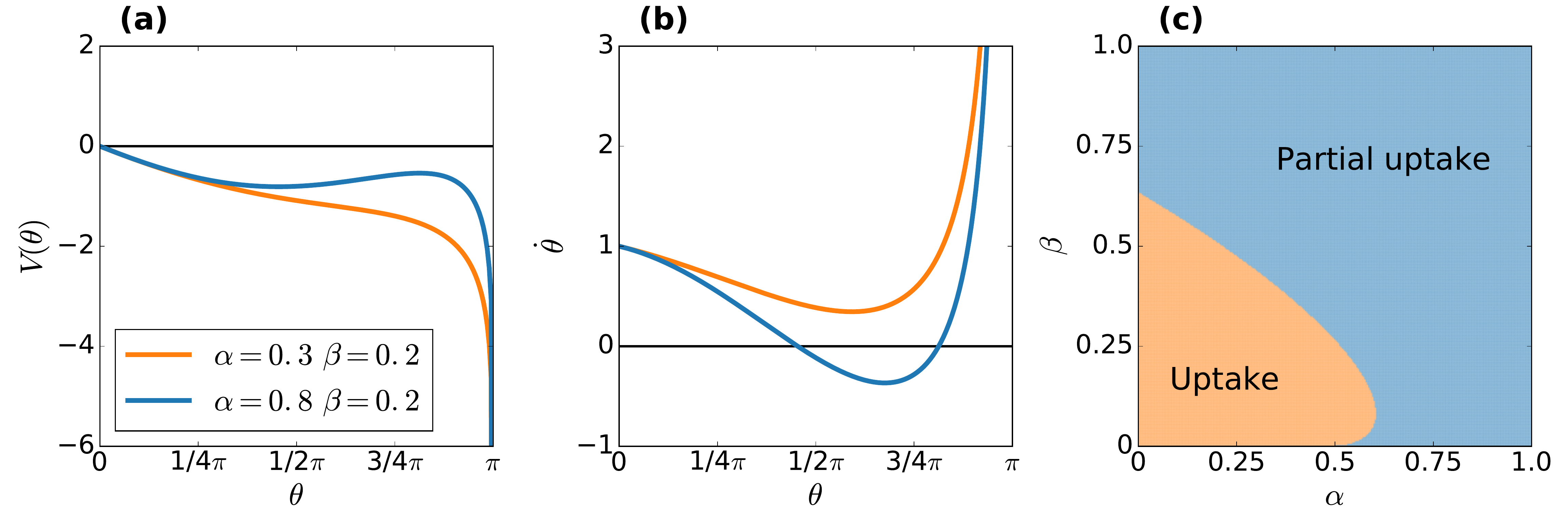}
\caption{
\label{fig:Figure13}
Deterministic uptake dynamics of spherical particles with the phenomenological contribution
of the free membrane.
(a) Uptake potential, leading to uptake (orange) and partial uptake (blue).
(b) Phase portrait ($\dot\theta$ vs.~$\theta$) for different parameter
values corresponding to uptake (orange) and partial uptake (blue). 
Parameters as in (a).
(c) Dynamical state diagram  
of the final uptake states as a function of $\beta$ vs.~$\alpha$.
Full uptake is achieved in the orange region and partial uptake in the blue region.
}
\end{figure*}

Interestingly, a moderate reduced line tension $\beta$ is productive as it 
increases the range of reduced membrane tension $\alpha$ for which full uptake
occurs. This effect however saturates
at $\beta=1$. Stronger values of $\beta$ are counter-productive as 
they transform partial uptake into no uptake. 
We note that for typical parameter values (using $\zeta=\sqrt{\kappa\sigma}$, i.e.~assuming
that the line tension is a result of the free membrane effects) 
one estimates  $\alpha=0.5$ and $\beta=0.5$, for which one would expect uptake.

In general, since the uptake dynamics depends on four parameters
($R$, $W$, $\kappa$  and $\sigma$) and $\alpha$ and $\beta$
are combinations of those, one would not move along horizontal or 
vertical lines in an experiment, where typically only one parameter is varied.
However, since $\beta/\alpha=\lambda/R$ one could move on
straight lines through the origin in the phase diagram when changing $W$ while
keeping $\lambda/R$ constant.

\subsection{Phenomenological description}

We now discuss the case when the simple line tension, Eq.~(\ref{line_tension}),
is replaced by the phenomenological description, Eq.~(\ref{free_mb_approx}), 
distilled out from the shape equations.
The additional factor of $\theta^2/2$ in Eq.~(\ref{free_mb_approx}) 
versus Eq.~(\ref{line_tension})
results in the dynamical equation
\begin{equation}
\frac{\mathrm{d} \theta}{\mathrm{d} \tau} =1 - 
\alpha (1-\cos \theta ) - \beta \theta \left(1+ \frac{1}{2} \theta \cot{\theta}\right) \, ,
\label{eq:det_ODE_ciricle_Two}
\end{equation}
with the same reduced membrane tension  
$\alpha$ and reduced line tension $\beta$ as before.
Now the last term cannot be integrated in closed analytical form as before.
%
%
The potential can be, however, easily obtained numerically as
displayed in Fig.~\ref{fig:Figure13}(a).
One clearly sees that the divergence at $\theta=0$, as occurring for the line tension, 
is now absent while the speed-up of uptake for large angle persists. 
Now only full uptake (orange) and partial uptake (blue) 
can be observed. However, we note that a trivial no uptake state
occurs for insufficient adhesion energy. 
Fig.~\ref{fig:Figure13}(b) shows the phase portrait and 
(c) the dynamical state diagram  for the uptake in the $\beta$-$\alpha$-plane.
Interestingly, the latter displays a re-entrance phenomenon at $\alpha$-values slightly above $1/2$,
meaning that upon increasing $\beta$ the system can
display partial uptake, full uptake and again partial uptake.
We note that the re-entrance phenomenon could be potentially 
observed when changing R while keeping $W$, $\sigma$ and $\kappa$ constant.

\begin{figure*}
\centering
\includegraphics[ width=1 \textwidth]{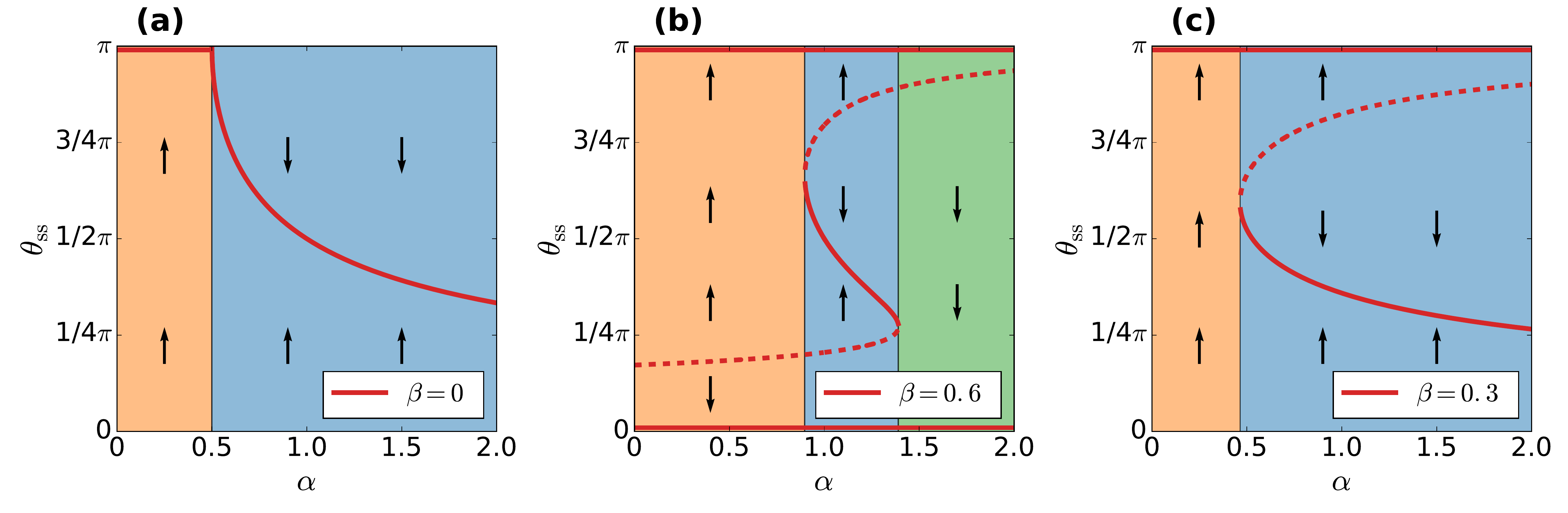}
\caption{Steady states as a function of the reduced membrane tension $\alpha$, 
in case of (a) no line tension ($\beta=0$),
(b) for a line tension with $\beta=0.6$ and
(c) including the phenomenological description of the free membrane, 
Eq.~(\ref{free_mb_approx}), for $\beta=0.3$.
Stable steady states are marked in solid, unstable ones as dashed.
The arrows display the flow of the system.
As usual, regions of full uptake are marked in orange, partial uptake in blue and no uptake in green.
Note that for (a) there is only one stable steady state and the dynamics is simple.
Line tension (b) and the free membrane effects (c) introduce new steady states
undergoing saddle-node bifurcations (see text).
}
\label{fig:Figure14}
\end{figure*}

Let us now compare and discuss the cases of
no line tension vs.~line tension vs.~the phenomenological treatment of the free membrane.
We will take a dynamical systems point of view, which turns out to be especially instructive.
Fig.~\ref{fig:Figure14} shows the steady states as a function of the 
reduced membrane tension $\alpha$, for the three cases.
As shown in Fig.~\ref{fig:Figure14}(a), without line tension ($\beta=0$)
the dynamics is quite simple: 
for $\alpha<1/2$ (i.e.~small reduced membrane tension), 
full uptake is achieved as $\theta=\pi$ is the only attractor.
As soon as $\alpha>1/2$  (i.e.~intermediate reduced membrane tension),
this only attractor moves to finite angles $\theta_{ss}<\pi$,
corresponding to partial uptake, decreasing further with increasing $\alpha$. 
The arrows mark the flow of the system, towards the stable attractors. Starting from the first adhesion
formed, corresponding to $\theta=0$, the system hence always evolves towards full or partial uptake,
depending on reduced membrane tension $\alpha$.

The reduced line tension, as visible in Fig.~\ref{fig:Figure14}(b), 
has two main effects: first, $\theta=0$ becomes a steady state,
due to the divergence of the potential. Second, another steady state emerges at intermediate angle, 
which has unstable (dashed) parts, but also a stable (solid) region corresponding to partial uptake. 
The bifurcation structure is the one of a pair of saddle-nodes.
While one assumes that the unstable branch for low reduced membrane tension 
$\alpha$ can be overcome by
the inherent fluctuations in receptor-ligand bond formation, hence still leading to full uptake, 
the unstable branch for large reduced membrane tension $\alpha$  
is at such a large angle that it should be interpreted as no possible uptake.
Finally, the stable branch in between the two saddle-nodes corresponds to the attractor
of partial uptake.  

\begin{figure*}
\centering
\includegraphics[ width=1 \textwidth]{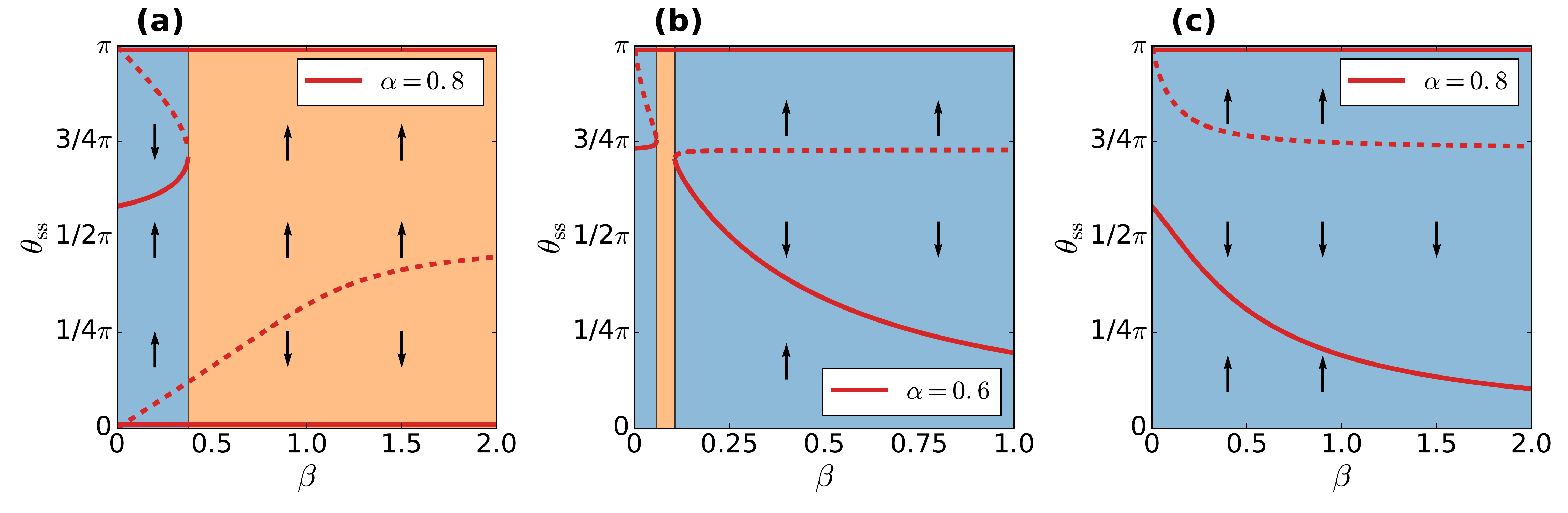}
\caption{
Steady states for line tension/free membrane effects as a function of 
reduced line tension $\beta$.
Stable steady states are marked in solid, unstable ones as dashed.
The arrows display the flow of the system.
As usual, regions of full uptake are marked in orange and partial 
uptake in blue. (a) Case of line tension and $\alpha=0.8$.
The partial uptake for small $\beta$ is due to the saddle-node,
whose stable branch prevents the system from reaching full uptake.
Increasing $\beta$ (and assuming the unstable branch can be overcome
by fluctuations) induces full uptake.
(b) Phenomenological treatment for the free membrane with $\alpha=0.6$.
Here, the system displays partial uptake, full uptake and again partial uptake, i.e.
a ``re-entrance'' of partial uptake occurs.
(c) Same as (b) but for $\alpha=0.8$ showing only partial uptake 
for increasing $\beta$.
}
\label{fig:Figure15}
\end{figure*}

Using the phenomenological description of the free membrane effects leads to 
the scenario shown in Fig.~\ref{fig:Figure14}(c). 
Here, the ambiguity of overcoming a barrier at small angle is absent, as $\theta=0$ 
is not a steady state anymore and no other branch prevents the flow from reaching $\theta=\pi$.
Only a single saddle-node emerges for larger reduced membrane tension
$\alpha$, separating uptake from partial uptake.
In a certain sense, the phenomenological description is an intermediate case
between having no line tension and a simple line tension.

Fig.~\ref{fig:Figure15} shows the dynamics and the dependence of the steady states
on the reduced line tension $\beta$, describing the strength of the line 
tension (a) or free membrane effects (b), (c), respectively.
One can clearly see that the emergence of saddle-nodes (one of whose branches has to be stable)
directly corresponds to the regions of partial uptake. 
Fig.~\ref{fig:Figure15}(b) shows an 
example of the re-entrance phenomenon, i.e.~where partial uptake emerges twice when 
varying the parameter $\beta$, cf.~also Fig.~\ref{fig:Figure13}(c).

To conclude, both a line tension and the approximated free membrane effects
considerably enrich the uptake dynamics.
They share similarities, for instance they accelerate uptake
as soon as the circumference of the membrane edge decreases, 
i.e.~in the second half of uptake ($\theta>\pi/2)$.
Although a partial uptake state can exist alone due to the 
interplay of adhesion energy and membrane tension that can generate 
a boundary minimum of the potential for spherical particles, 
here another partial uptake state can emerge.
This partial uptake state is caused by a non-boundary 
minimum of the potential due to the saddle-node structures of the new 
steady states that is generated either by line tension or 
approximated free membrane effects. While the line tension can introduce
no uptake states that are caused if a repellor at large angle 
values emerges, no uptake states only emerge
in the phenomenological description in the trivial case of insufficient
adhesion energy.
The main difference between them, however, is that the line tension 
introduces an energy barrier at small angles,
that has to be overcome by fluctuations.
Therefore a line tension is not an adequate description for the free membrane, 
which does not show this effect. 
We stress that this conclusion 
remains true beyond the approximation used in Eq.~(\ref{free_mb_approx}): 
Eq.~(\ref{eq:Foret_loose}) leads to
$E_{\mathrm{free}}^{\mathrm{tot}}\propto  \sin^4 \theta$
for small angles, displaying the same behavior, i.e.~no barrier.

\section{Stochastic dynamics}
\label{sec:six}

\subsection{General approach}

\begin{figure}
\centering
\includegraphics[ width=.48 \textwidth]{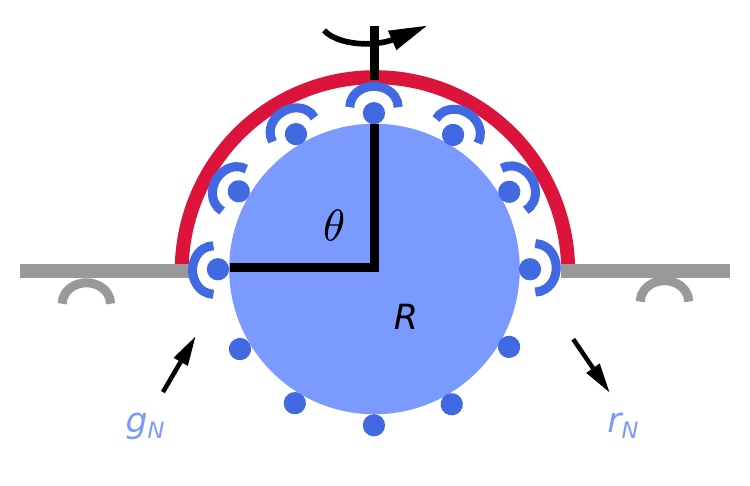}
\caption{ 
Modeling particle uptake as a discrete stochastic process. 
The particle (blue) is covered with ligands (small discs) that stochastically bind 
to cell surface receptors (half-circles) with rate $g_N$, leading to an advancement of the
adhered membrane area, or unbind with rate $r_N$. Note that axial symmetry is assumed.
}
\label{fig:Figure16}
\end{figure}

For the uptake of both nanoparticles or viruses, fluctuations are 
expected to be important since the particles are small and 
typically only covered by few tens of ligands,
rendering ligand-receptor binding a discrete stochastic process. 
Based on the continuous deterministic modeling approach, 
we now explicitly model the discrete stochastic dynamics of 
receptor-ligand binding as sketched schematically in Fig.~\ref{fig:Figure16}.
Our stochastic model is defined from the deterministic one
via two steps. First, the system is discretized by mapping the continuous deterministic
equation onto a discretized version.
Secondly, this equation is interpreted as a stochastic rate equation.
The considered stochastic process is designed to 
obtain the deterministic result in the limit of small noise, i.e.~in the
continuous deterministic limit. However, we do not introduce 
temperature and our stochastic model is not designed to 
obtain the deterministic result in the limit of vanishing temperature.
The underlying reason is that we do not want to make
assumptions about the processes that drive the membrane
forward and backward. Modelling the growth of adhered membrane
patches based on receptor-ligand binding is a mature and
challenging field by itself \cite{reister2008dynamics,weikl2009adhesion} and
might even include active processes \cite{turlier2016equilibrium}.
Therefore we do not anchor our model in an equilibrium model,
but derive our stochastic rates from the deterministic theory
without enforcing detailed balance, similar to stochastic processes
used in population or evolutionary dynamics \cite{cremer2012growth,melbinger2010evolutionary}.

In detail, we first map the adhered 
membrane area onto the number of bound ligands $N$, 
in order to deduce a discrete differential equation
$\mathrm{d} N / \mathrm{d} t=\mathrm{d} N / \mathrm{d} x\cdot \mathrm{d} x /\mathrm{d} t $, 
where $x=\{z,\theta \}$  corresponds to the uptake variable (height $z$
for the cylinder in normal orientation and invagination angle 
$\theta$ for the parallel cylinder and the sphere) \cite{frey2019}.
From the deterministic framework, $ \mathrm{d} x /\mathrm{d} t$ is known from 
Eq.~(\ref{eq:det_dyn_NC}), Eq.~(\ref{eq:det_dyn_PC}) and Eq.~(\ref{eq:det_ODE})
for normal cylinders, parallel cylinders and spheres, respectively.
The next step is to deduce the corresponding one-step Master equation (ME) \cite{vankampen1992}  
for the probability $p_N$ to have $N$ ligands bound to receptors, 
\begin{equation}
\frac{\mathrm{d} p_N}{\mathrm{d} \tau} = 
g_{N-1} p_{N-1}+r_{N+1} p_{N+1} -(g_N+r_N)p_N.
\end{equation}
Here $g_N$ is the forwards and $r_N$ the backwards rate
by which ligands bind (unbind) from state $N$. 
While complete probability distributions can be calculated analytically only
for some special cases, the stochastic uptake times $T_\mathrm{sto}$ for 
a reflecting boundary at the unwrapped state and an absorbing
boundary at the fully wrapped state can be calculated analytically as \cite{vankampen1992}
\begin{equation}
T_\mathrm{sto}=\sum_{\nu=1}^{N_\mathrm{max}-1} \sum_{\mu=1}^{\nu}
\frac{r_\nu r_{\nu-1} \cdot \cdot \cdot r_{\mu+1}}
{g_\nu g_{\nu-1} \cdot \cdot \cdot g_{\mu}} \, .
\label{eq:analytical_uptake_times}
\end{equation}
Alternatively, one can use the Gillespie algorithm \cite{gillespie1977} to
simulate both the probability distributions and the uptake times.

\subsection{Cylinder with normal orientation $(\bot)$}

For a cylinder oriented perpendicularly to the membrane,
the membrane covered area $A_\mathrm{ad}^\bot = A(z)$ is mapped
onto the number of bound receptors $N^{\bot}(z)$ 
by using 
\begin{equation}
A(z)/A_{\mathrm{max}}^{\bot} = 
(N^{\bot} (z)-1)/(N_\mathrm{max}-1).
\end{equation}
Here we assumed that initially, the particle
is already bound to the membrane by one ligand, 
yielding $N^{\bot} (z) = (N_\mathrm{max} - 1) z / L +1$.
The corresponding discrete equation then reads
\begin{equation}
\frac{ \mathrm{d} N^{\bot} }{\mathrm{d} t} = 
\frac{N_\mathrm{max}-1}{L} \left( \nu_{\mathrm{w}}^{\bot}-
\nu_{\mathrm{\kappa}}^{\bot}-\nu_{\mathrm{\sigma}}^{\bot} \right ) \, ,
\end{equation}
and the
corresponding rates of the ME are hence easily deduced by
$g_N=(N_\mathrm{max}-1)\nu_{\mathrm{w}}^{\bot}/L$ and
$r_N=(N_\mathrm{max}-1)(\nu_{\mathrm{\kappa}}^{\bot}+
\nu_{\mathrm{\sigma}}^{\bot})/L$.
Finally, to implement a reflecting boundary condition at $N=1$,
we put $r_1$=0.

\subsection{Cylinder with parallel orientation $(\parallel)$}

In this case we proceed 
similarly and map the membrane covered area 
$A_\mathrm{ad}^{\parallel} = A(\theta)$ 
onto the number of bound receptors $N^{\parallel}  (\theta)$
via
$ A(\theta)/A_{\mathrm{max}}^\parallel = (N^{\parallel}  (\theta)-1)/
(N_\mathrm{max}-1)$. One intitially bound ligand implies
$N^{\parallel}  (\theta) = (N_\mathrm{max} - 1) \theta /\pi+1$.
The corresponding discrete equation reads
\begin{equation}
\frac{ \mathrm{d} N^{\parallel}  }{\mathrm{d} t} = 
\frac{N_\mathrm{max}-1}{\pi} \left( \nu_{\mathrm{w}}^{\parallel}-
\nu_{\mathrm{\kappa}}^{\parallel}-\nu_{\mathrm{\sigma}}^{\parallel} 
(1-\cos \theta )
 \right ) \, ,
\end{equation}
and hence the corresponding rates of the ME amount to 
$g_N=(N_\mathrm{max}-1)\nu_{\mathrm{w}}^{\parallel}/\pi$ and
$r_N=(N_\mathrm{max}-1)(\nu_{\mathrm{\kappa}}^{\parallel}+
\nu_{\mathrm{\sigma}}^{\parallel}(1-\cos \theta))/\pi$. 
Finally, to implement a reflecting boundary condition at $N=1$,
we again put $r_1$=0.

\subsection{Sphere $(\circ)$}

Finally, for a spherical particle we map 
$A_\mathrm{ad}^{\circ} = A(z)$ onto $N^{\circ}(\theta)$ by
\begin{equation}
A(\theta) / A_{\mathrm{max}}^{\circ} =  (N^{\circ}(\theta)-1)/
(N_{\mathrm{max}}-1) \, ,
\end{equation}
and one initially bound ligand implies
$ N^\circ =(N_{\mathrm{max}}-1)(1-\cos(\theta) ) /2+1$.
The discrete equation reads
\begin{equation}
\frac{ \mathrm{d} N^{\circ}  }{\mathrm{d} t} = 
\left ( \nu_{\mathrm{w}}^\circ-\nu_{\kappa}^\circ-\nu_{\sigma}^\circ
 (1-\cos \theta )  - \nu_{\zeta}^\circ \cot{\theta}  \right ) N_{\mathcal{E}}(N).
 \label{eq:det_discrete_ODE}
\end{equation}
Here $N_{\mathcal{E}}(N)=\sqrt{(N-1)((N_{\mathrm{max}}-1)-(N-1))}$
is the number of receptors at the advancing edge.
The corresponding rates of the ME are given by
$g_N=\nu_{\mathrm{w}}^{\circ}  N_{\mathcal{E}}(N)$
and
$r_N= ( \nu_{\kappa}^\circ + \nu_{\sigma}^\circ
 (1-\cos \theta )  + \nu_{\zeta}^\circ \cot{\theta}  )   N_{\mathcal{E}}(N)$
for $\theta< \pi/2$ and
$g_N=(\nu_{\mathrm{w}}^{\circ}  - \nu_{\zeta}^\circ \cot{\theta})N_{\mathcal{E}}(N)$
and
$r_N= \left ( \nu_{\kappa}^\circ + \nu_{\sigma}^\circ
 (1-\cos \theta ) \right)   N_{\mathcal{E}}(N)$
for $\theta\ge\pi/2$ in order to ensure that the rates stay positive.
In case of the phenomenological treatment of the free membrane parts,
the last term in the bracket of $r_N$ and $g_N$
is replaced by $ \nu_{\zeta}^\circ \, \theta (1 +\theta/2 \cot{\theta})$
in the intervals $\theta <2.289$ and $\theta \ge 2.289$, accordingly.

Finally, note that in case of the sphere one has to make additional assumptions for the rate $g_1$,
which otherwise would be zero. 
We here choose  $g_1=\nu_{\mathrm{w}}^{\circ}  \sqrt{N_\mathrm{max}}$, since
(i) it should be proportional to $\nu_w$ and (ii) 
the transition time from state $N=1$ to state $N=2$ 
should vanish for large $N_\mathrm{max}$.
Since $r_1=0$, $N=1$ is a reflecting pure boundary, 
i.e.~a particle always stays attached to the membrane.

\begin{table}[b]
\caption{Simulation parameters.}
\centering
\begin{ruledtabular}
\begin{tabular}{ccc}
Parameter & Used value & Ref. \\
\hline
Bending rigidity &$\kappa=25 \,\mathrm{k_{B}} T$ & \cite{kumar2016} \\
Membrane tension &$\sigma=1\cdot 10^{-5}\, \rm N/m$& \cite{foret2014}\\
Energy density &$W=0.04\, \rm{mJ / m^{2}}$ &\cite{agudo2015}\\
Membrane microviscosity & $\eta=1\, \rm Pa \;s$& \cite{agudo2015} \\
Line tension &$\zeta=\sqrt{\kappa \sigma}$& \\
Receptor-ligand pairs &$N_{\mathrm{max}}=20$ &   \\
\end{tabular}
\label{tab:SimulationParameters}
\end{ruledtabular}
\end{table}

\begin{figure}
\centering
\includegraphics[ width=0.47 \textwidth]{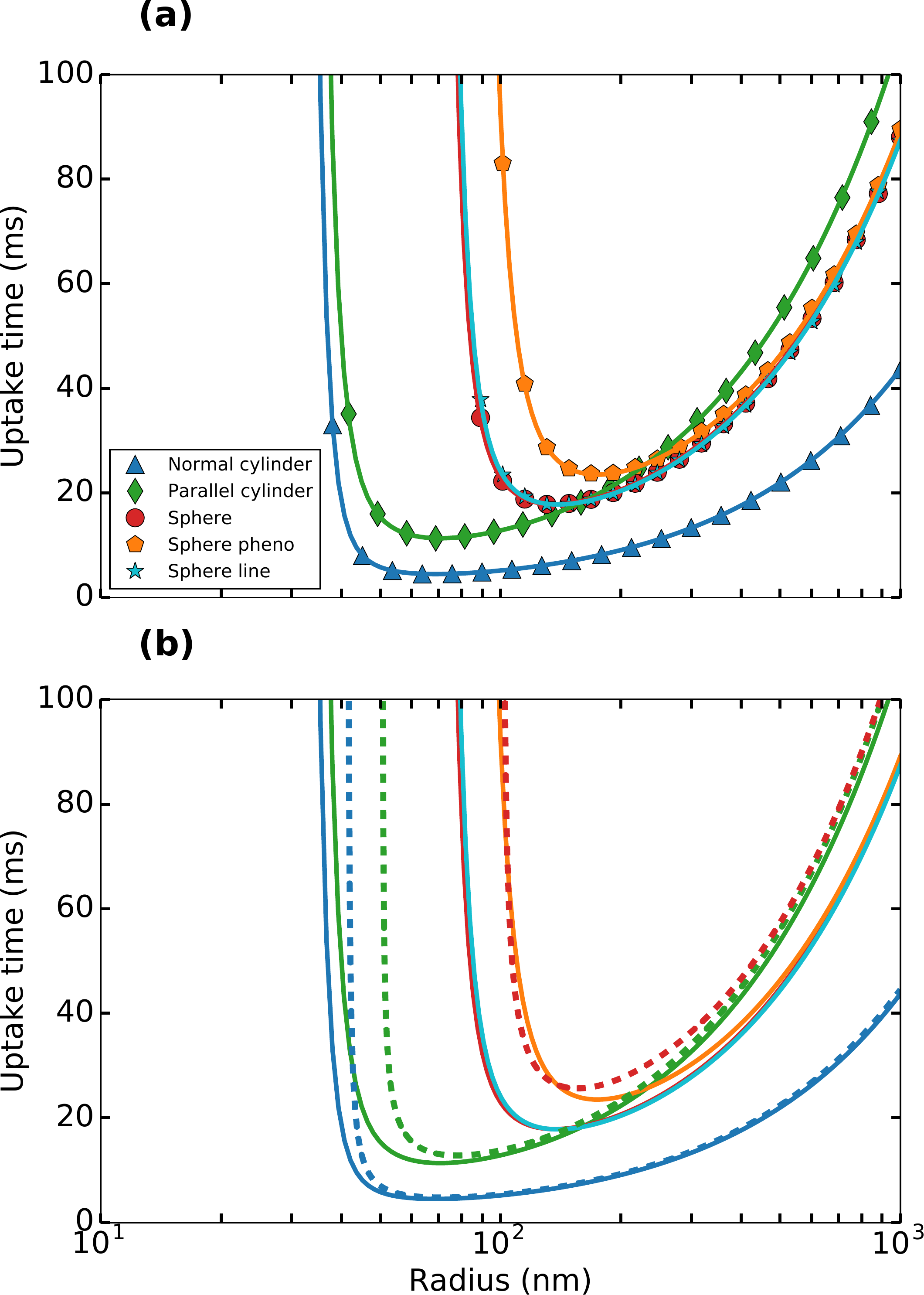}
\caption{
(a) Simulated mean uptake times (symbols) for a 
normally oriented cylinder (blue triangles), 
a parallelly oriented cylinder (green lozenges), a sphere (red circles),
a sphere including the phenomenological approximation for the free membrane 
(orange pentagons) and a sphere with line tension (cyan stars)
as function of radius at equal volume and radius for the parameter values in 
Table~\ref{tab:SimulationParameters}.
In addition, the numerically computed mean uptake times calculated by
Eq.~(\ref{eq:analytical_uptake_times}) are shown as solid lines for the different
shapes in the corresponding colors.
(b) For the normal cylinder, the parallel cylinder and the sphere the deterministic uptake 
times are shown as dashed lines in corresponding colors (for vanishing line 
tension/free membrane) and compared to the numerically calculated mean uptake
times (solid lines).
}
\label{fig:Figure17}
\end{figure}

\subsection{Stochastic uptake times}

We numerically solve the MEs corresponding to the different particle 
shapes and orientations by means of the Gillespie algorithm \cite{gillespie1977}. 
The used parameter values are summarized in 
Table~\ref{tab:SimulationParameters}.
For the number of ligands we chose a typical value of $N_\mathrm{max}=20$.
Stochastic effects further increase upon decreasing $N_\mathrm{max}$, and they prevail
for $N_\mathrm{max}$ of the order of one hundred \cite{frey2019},
depending on parameters.

In order to obtain the stochastic uptake times, we introduce an 
absorbing boundary at full coverage.
Fig.~\ref{fig:Figure17}(a) shows the uptake time as a function
of the radius of the particle. The results of our simulations (symbols), 
averaged over $10^4$ trajectories each, perfectly agree with the results 
of the numerical calculations by means of
Eq.~(\ref{eq:analytical_uptake_times}) (solid lines), which verifies our simulations.
To compare cylinders to spheres,
their radius were fixed to the one of the respective sphere 
and the length $L$ was adjusted to obtain equal volume. 
Shown in the figure are the mean uptake times obtained by 
stochastic simulations (averaged over $10^4$ trajectories each) 
for the normally oriented cylinder (blue triangles), the parallelly oriented cylinder (green lozenges), 
the sphere (red circles), the sphere including the phenomenological treatment 
of the free membrane,
according to Eq.~(\ref{eq:scaling}) and Eq.~(\ref{free_mb_approx}) 
(orange pentagons) and the sphere including a line tension (cyan stars),
according to Eq.~(\ref{line_tension}) and Eq.~(\ref{eq:scaling}).

In Fig.~\ref{fig:Figure17}(b) the mean uptake times obtained by 
numerical calculations for the different particle shapes are compared to the
deterministic results (without line tension or free membrane effects), which are 
shown as the dashed curves in the corresponding colors.
We see that for all considered particle shapes both the deterministic and the stochastic dynamics 
show similar behavior. 
First, a critical radius exists below which uptake is not possible anymore 
(cf.~the analytical results in section \ref{sec:three} B,C,D) \cite{LipoDoeb}.
Second, for larger radii the uptake time increases with
increasing particle size. 
And third, in between an optimal radius exists, having minimal uptake time 
\cite{osaki2004quantum,gao2005,zhang2009size,huang2013role}.
The underlying reason for this behavior is that the bending energy is independent
of particle size, while the energy contributions for adhesion and tension increase with size
\cite{LipoDoeb}. 

\begin{figure*}
\centering
\includegraphics[ width=1. \textwidth]{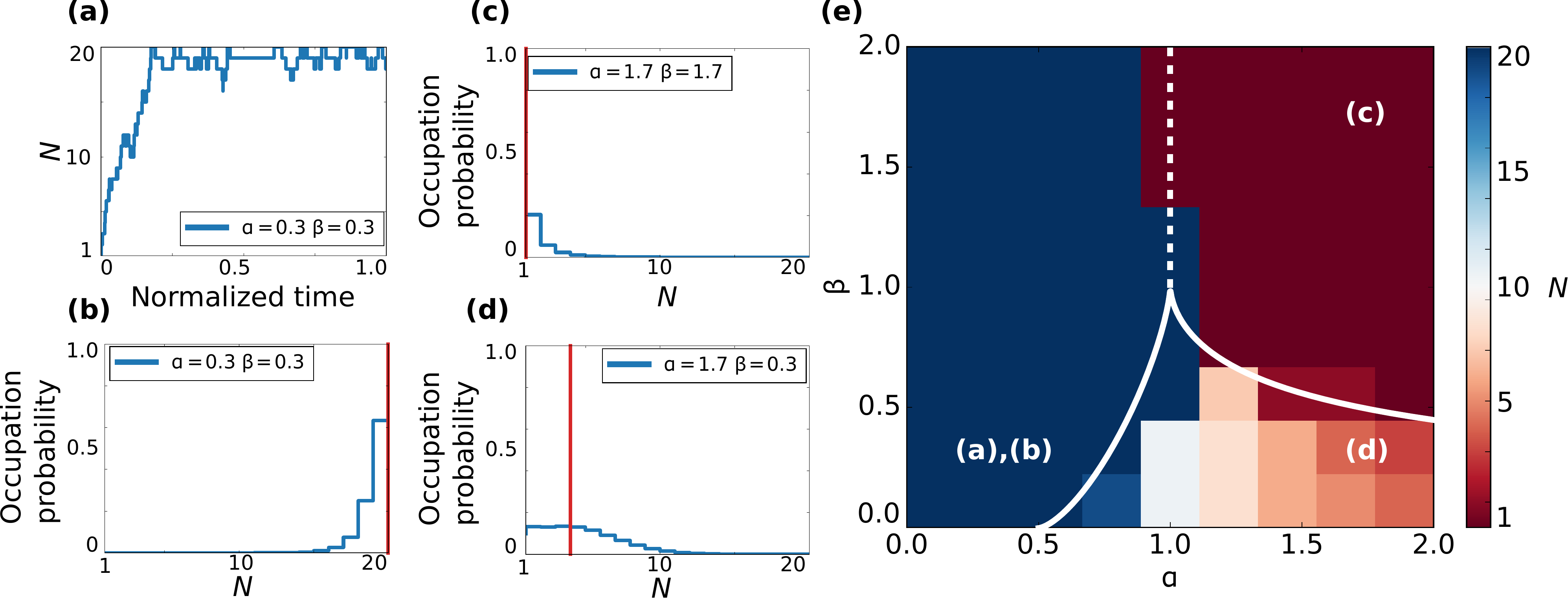}
\caption{
Stochastic uptake dynamics of spherical particles with line tension.
(a) Uptake trajectory of a spherical particle with $\alpha=0.3$
and $\beta=0.3$ for $N=10^2$ time steps.
(b)-(d) Occupation probability during uptake with 
$N_\mathrm{max}=20$ receptors for
different parameter combinations for an uptake trajectory of $N=10^6$
time steps. The receptor with largest occupation probability is shown in 
red. (b) Uptake. (c) No uptake. (d) Partial uptake.
(e) Dynamical state diagram of stochastic uptake showing the maximum
occupation probability of a trajectory of $N=10^6$ time steps.
The white lines correspond to the state boundaries of
the deterministic model, cf. Fig.~\ref{fig:Figure12}(c).}
\label{fig:Figure18}
\end{figure*} 

From Fig.~\ref{fig:Figure17} we further see that
stochastic uptake is always faster than deterministic uptake. 
In addition, stochasticity extends uptake towards regions beyond 
the critical radius of deterministic uptake.
This is due to fluctuations being able to drive particles 
above energy barriers during uptake. 
Another interesting observation is that in the deterministic description, 
cylinders in the parallel orientation
are taken up faster than (or at least equally fast as) spherical particles for all radii. 
In contrast, the stochastic description states that parallelly oriented cylinders 
are only faster than spheres below a certain radius: for the given parameters, 
at around $R=150\,{\rm nm}$, 
the situation reverses and spheres are taken up faster \cite{frey2019}.
In addition, we note that the typical time scale for uptake is between a
few ms to a few tens of ms, which is in agreement with previous simulation
results \cite{huang2013role}.

We now discuss the effects of the line tension and the free membrane.
As can be judged from Fig.~\ref{fig:Figure17}(a) when comparing 
red circles, orange pentagons and cyan stars, 
when a line tension is used to model the free membrane parts \cite{LipoDoeb}, suggesting
the scaling $\zeta=\sqrt{\kappa\sigma}$ \cite{DesernoPRE04},
or when incorporating the phenomenological approximation 
given by Eq.~(\ref{eq:scaling}) and
 Eq.~(\ref{free_mb_approx}), the effects are relatively modest
in regard to the uptake times. 
One nevertheless can see that the line tension slightly hinders uptake, 
while the phenomenological treatment slows down the uptake of small particles
even more.
For increasing particle sizes the effect vanishes as $\nu_\zeta \sim 1/R^2$.
Furthermore, as the effect of the free membrane only slightly affects
uptake times and contributes up to $20\%$ to the total deformation 
energy of the membrane, neglecting the free membrane 
as done in Ref. \cite{frey2019} is justified in regard to uptake 
times and for typical parameters for nanoparticle and virus uptake.
However, a line tension can also originate from a localization of lipids or curvature
generating proteins at the border between adhered and free membrane \cite{lipowsky1992budding}.
Then $\zeta$ can possibly be larger and hence the effect of the slowing down of the uptake
would be more pronounced.

\subsection{Stochastic state diagrams}

We finally make contact between our stochastic model and the
state diagrams for the deterministic dynamics presented above.
Fig.~\ref{fig:Figure18} shows the stochastic uptake dynamics of 
spherical particles with line tension with $N_\mathrm{max}=20$
ligand-receptor pairs.
As we here simulated the non-dimensionalized 
Eq.~(\ref{eq:det_ODE_ciricle}) 
we use the following rates $g_N=N_\mathcal{E}$
and $r_N=(\alpha(1-\cos \theta)+\beta \cot \theta )N_\mathcal{E}$
for $\theta < \pi/2$
and
$g_N=(1-\beta \cot \theta )N_\mathcal{E}$
and
$r_N=\alpha(1-\cos \theta)N_\mathcal{E}$
for $\theta \ge \pi/2$.
In addition, we now implement reflecting boundary
conditions both for $\theta=0$ by $g_1=\sqrt{N_\mathrm{max}}$, $r_1=0$, 
and $\theta=\pi$ by $g_\mathrm{max}=0$,
$r_\mathrm{max}=\alpha(1-\cos \theta) \sqrt{N_\mathrm{max}}$
to study the occupation probabilities of different states for long times.

Fig.~\ref{fig:Figure18}(a) shows an example of an uptake trajectory for $N=10^2$ time steps.
The parameters are chosen such that uptake is expected.
From uptake trajectories of $N=10^6$ time steps
 the occupation probabilities of the different states are computed
(b)-(d). To classify the uptake state we calculate the state of the largest
occupation probability which is shown by the red vertical line.
We have uptake when the $N=20$ state has the largest
probability (b). Similarly, we find no uptake if the
N = 1 state has the largest occupation probability (c).
Highest probability for any other state corresponds to partial uptake (d).
Using this classification we calculated the dynamical state diagram of 
stochastic uptake of trajectories of $N=10^6$ time steps in the
$\beta$-$\alpha$-plane (e).
The white lines correspond to the state boundaries of Fig.~\ref{fig:Figure12}(c)
and we see surprisingly similar behaviour.
Compared to Fig.~\ref{fig:Figure12}(c), we find that the parameter region 
where uptake is possible is slightly extended to the partial uptake region. 
In addition, we confirmed that, as assumed above, 
fluctuations allow the system to cross the initial barrier caused 
by the line tension as discussed in section~\ref{sec:five}.
In order to determine the maximum angle which can be 
overcome by fluctuations we used the fact that uptake is possible 
in Fig.~\ref{fig:Figure18}(a) as long as $\alpha\le1$. For $\alpha=1$ 
the root of Eq.~(\ref{eq:det_ODE_ciricle}) is given by $\pi/2$
that is the maximum angle, independent of $\beta$.

\begin{figure}
\centering
\includegraphics[ width=0.48 \textwidth]{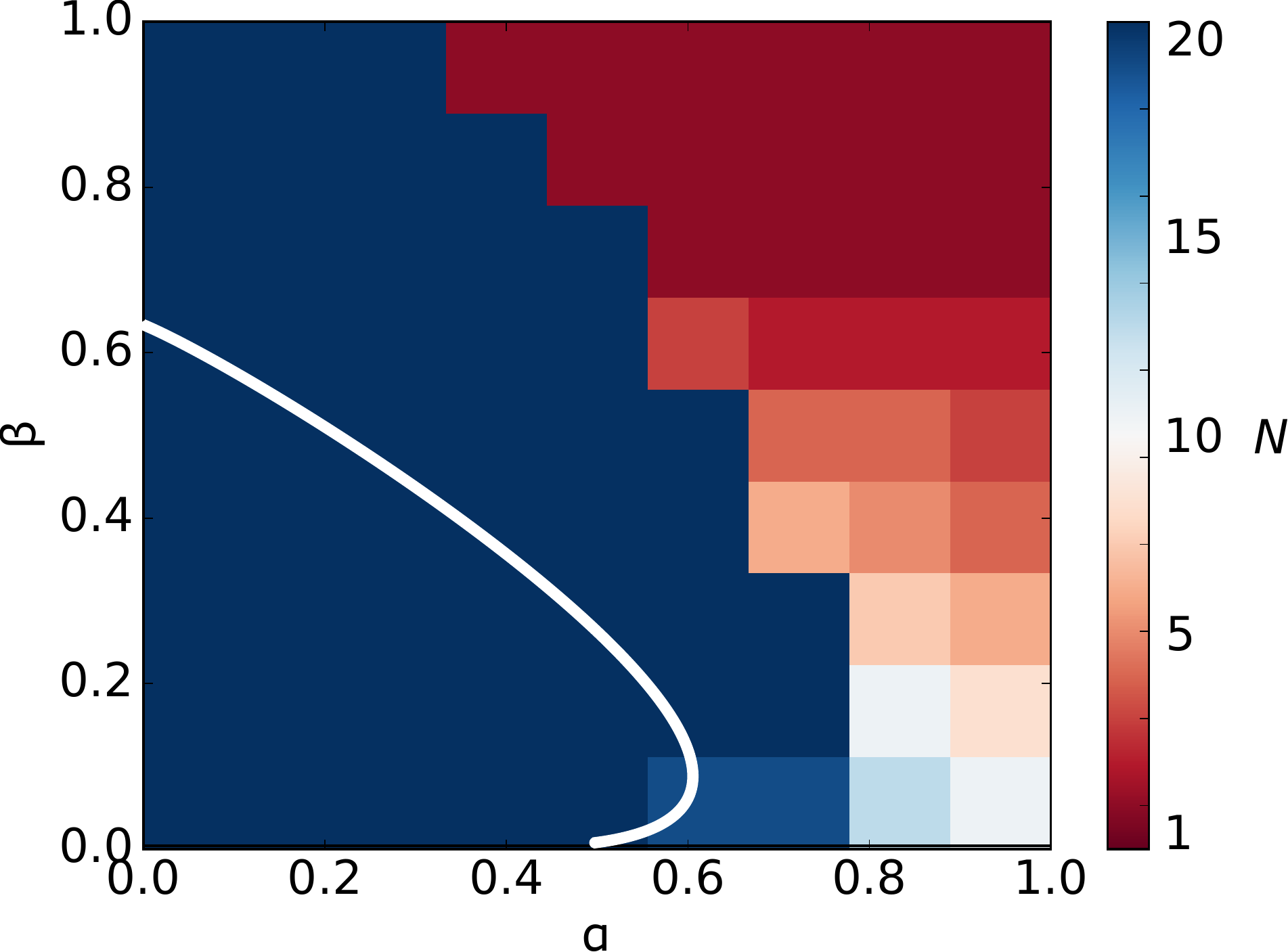}
\caption{
Dynamical state diagram of stochastic uptake 
of spherical particles using the phenomenological description
of free membrane effects 
showing the maximum
occupation probability of a trajectory of $N=10^6$ time steps.
The white line corresponds to the state boundary of
the deterministic model, cf. Fig.~\ref{fig:Figure13}(c).
}
\label{fig:Figure19}
\end{figure}

Finally, we study stochastic uptake where we include the free membrane effects
by our phenomenological description. We now 
analyze the non-dimensionalized  Eq.~(\ref{eq:det_ODE_ciricle_Two}).
In this case the forward and backward rate change to
$g_N=N_\mathcal{E}$,
$r_N=(\alpha(1-\cos \theta)+\beta \theta (1+\theta/2 \cot \theta ))N_\mathcal{E}$ 
for $\theta < 2.289$ and 
$g_N=(1-\beta \theta (1+\theta/2 \cot \theta ))N_\mathcal{E}$, 
$r_N=\alpha(1-\cos \theta)N_\mathcal{E}$ for $\theta\ge2.289$,
whereas all other rates stay identical.
In Fig.~\ref{fig:Figure19} we calculated the dynamical state 
diagram of stochastic uptake of trajectories of $N=10^6$ time steps in the
$\beta$-$\alpha$-plane using the same classification
procedure as before.
Compared to Fig.~\ref{fig:Figure13}(c) we find that now uptake also
extends beyond the state boundaries of the deterministic calculation (white line),
because fluctuations drive the system above the intermediate barrier, corresponding
to a partial uptake state. Without fluctuations the partial uptake state could not be 
overcome in this parameter region. 
In addition, we find that similar to Fig.~\ref{fig:Figure13}(c)
either uptake or partial uptake occurs.

\section{Conclusion}
\label{sec:seven}

Here we have studied the uptake dynamics of nanometer-sized
particles such as viruses at cell membranes mediated by ligand-receptor binding.
The focus was on a simple but complete deterministic model, amenable to analytical insight,
complemented by stochastic simulations.

By considering the adhesion, the bending and the tension energies of the cell
membrane, including the contributions of the free, i.e.~non-adhered membrane parts
and calculating minimal energy membrane shapes numerically,
we found that in the parameter regime which is typically relevant for biological systems, 
the free membrane contributes up to $20\%$ to the total energy 
during uptake. It thus cannot be neglected \textit{a priori}.
We hence incorporated the free membrane effects within a simple dynamical model for uptake
by either a line tension or by an effective phenomenological description.
This allowed us to study the deterministic uptake dynamics for spherical,
cylindrical and spherocylindrical particles (both oriented either perpendicular 
or parallel to the membrane).

Similar to \cite{huang2013role,dasgupta2014}, where the uptake of 
spherocylindrical particles was studied, we find that the aspect ratio of cylindrical
particles dictates the uptake pathway.
While short cylinders are taken up fastest in normal mode, long cylinders
are taken up fastest in parallel mode.
For long cylinders at large reduced membrane tension one could speculate
that they are taken up initially in parallel mode but might reorient driven by
fluctuations to the perpendicular position in order to complete the uptake process
\cite{huang2013role}.
Calculating the uptake times, spherical particles were 
found to be always taken up slower than cylinders.
For spherocylindrical particles we found that they are always
taken up fastest in parallel mode at equal volume and aspect ratio.
When comparing them to spheres we found that while
normal spherocylinders are taken up slower, for parallel spherocylinder
the result depends on aspect ratio at equal volume.
Our results regarding cylinders might change if
one would include detailed models for adhesion and membrane
shapes at the bottom and top faces. Here these effects
are neglected to keep the model transparent.

For spherical particles, the free membrane effects influence the dynamics. 
In accordance with earlier work \cite{deserno2002adhesion,
DesernoPRE04,vacha2011receptor,yi2011cellular,dasgupta2014}, 
we could identify in the analytical model three scenarios for both the 
line tension and the phenomenological description
of the free membrane: full uptake, partial uptake and no uptake,
dictated by membrane elasticity, adhesion energy and the free membrane.
There are, however, differences: the line tension induces an energy barrier 
(when considering the total energy) for small uptake angles for all parameters,
hence uptake is only possible if assisted by fluctuations.
In contrast, for the phenomenological description, the existence of this barrier 
depends on the parameters $\alpha$ and $\beta$.
This is similar to the earlier work by Deserno and Gelbart 
\cite{deserno2002adhesion}, 
where the shape of the free membrane shape is represented by a torus, 
that prevents uptake depending on particle size and the elasticity of the 
membrane.
Considering the uptake speed, both the line tension and the 
phenomenological description speed up the process when
passing over the equator. Overall
we conclude that a line tension is not the best approach to describe the
effects of the free membrane in the intermediate regime between tense
and loose membranes. The phenomenological approach suggested here
is clearly a better approximation. We presented a complete
analysis of this case, focusing on steady states and dynamical state diagrams
as a function of a reduced membrane tension ($\alpha$)
and a reduced line tension ($\beta$). The occurrence of 
parameter regions of partial/no uptake
could be traced back to new steady states emerging via saddle-node bifurcations.

Finally, we included stochasticity into the model as receptor-ligand binding is a
discrete process in a small system such that fluctuations are expected to be important. 
This was achieved by mapping the deterministic models onto one-step master equations.
In a first step, we used these to simulate and calculate uptake times. 
We could show that the effect of spheres profiting from noise and getting
taken up faster than parallel cylinders, as described recently \cite{frey2019}, 
survives when free membrane effects are included. In a second step, 
we calculated stochastic state diagrams and again found surprisingly
good agreement with the deterministic results. In both cases, it became clear
that stochasticity enlarges the parameter region in which uptake is possible.
Thus, fluctuations are expected to help the 
system to transverse barriers, corresponding to no uptake or 
partial uptake states, which
could not be overcome in a deterministic model. In the future, our model
could be extended by more detailed assumptions regarding the 
way the membrane moves forward and backward at the contact line.
Such a description then could also describe the effect of temperature, which
is not treated here, and of active membrane fluctuations, which would
break detailed balance again.

Our uptake times are a lower bound to experimentally measured
uptake times because we only consider the dissipative forces resulting
from membrane microviscosity. 
Other mechanisms that might be rate-limiting include receptor diffusion within 
the plasma membrane, yielding uptake times in the few seconds 
range \cite{gao2005},
the assembly of clathrin lattices, with typical uptake times of the order
of 60 seconds \cite{kaksonen2018mechanisms},
dissection of the cytoskeleton beneath the plasma membrane \cite{stolp2011hiv},
and scission, with a timescale of minutes \cite{shlomovitz2011membrane}.
However, there is an increasing amount of experimental results 
showing that uptake times can indeed occur on the timescale of
tens of milliseconds, including 250 ms 
for human enterovirus 71 \cite{pan2017process},
around $\unit[80]{ms}$ for gold nanoparticles ($\unit[20]{nm}$) 
\cite{ding2015recording},
below $\unit[40]{ms}$ for the uptake of micrometer-sized 
latex beads by GUVs \cite{dietrich1997adhesion}
and even below 20 ms for silicon nanoparticles of ($\unit[4]{nm}$) 
diameter \cite{wang2019monitoring}. More importantly, however, is
that our theory predicts very interesting effects regarding relative uptake
times for different shapes. Our predictions in regard to the
relative sequence of uptake and the phase diagrams do not
depend on the absolute scale of the uptake time. They would only
change if the dissipative uptake dynamics were not dominated by
one single local time scale, for example if the advancement of the 
adhesion front was strongly limited by receptor diffusion \cite{gao2005,boulbitch2001kinetics}.
Finally our work suggests that both during biological evolution
and for particle design in materials science, stochasticity might play an important role for 
optimal performance, which here is identified with fast uptake.

\begin{acknowledgments}
F.F. acknowledges support by the Heidelberg Graduate
School for Fundamental Physics (HGSFP). We acknowledge the members of the
Collaborative Research Centre 1129 for stimulating discussions on viruses. 
U.S.S. acknowledges support as a member of the Interdisciplinary
Center for Scientific Computing (IWR) and the clusters of excellence CellNetworks,
Structures and 3DMM2O.
\end{acknowledgments}

\appendix

\section{Details on solving the shape equations}
\label{appendix:details_shapeeq}

To numerically solve the boundary value problem 
given by Eq.~(\ref{eq:shape}), we rewrite it as a system of four first order
ordinary differential equations. 
Three boundary conditions in Eq.~(\ref{eq:boundaryconditions}) are given for 
$s\rightarrow \infty$. We hence use the asymptotic solution 
to shift the boundary conditions for the numerical problem to a finite arc length $s_\mathrm{max}$. 
For weak membrane deformations  ($\psi\ll1$) the linearized shape equations are solved by \cite{foret2014}
\begin{equation}
r(s)=s \, , \;\;\; \psi(s) =\beta K_1(s/\lambda) \, , \;\;\;  z(s)=\beta \lambda K_0(s/\lambda) \, ,
\end{equation}
where $\beta=\theta /  K_1 (R/\lambda \sin \theta ) $ is a parameter and 
$K_n$ are the modified Bessel functions of the second kind.
Then the numerical boundary conditions using matched asymptotics 
read
\begin{align}
r(0)&=R \sin(\theta),
\nonumber \\
\psi (0)&=\theta,
\nonumber \\
\psi(s_\mathrm{max}) &=\beta {K_1 (s_\mathrm{max}/\lambda)}, 
\nonumber \\ 
\dot{\psi}(s_\mathrm{max}) &=-\frac{\beta}{2 \lambda} \left 
( K_0(s_\mathrm{max}/\lambda)+K_2(s_\mathrm{max}/\lambda) \right ), 
\nonumber \\ 
z(s_\mathrm{max})&=\beta \lambda  K_0(s_\mathrm{max}/\lambda).
\end{align}
The matching point $s_\mathrm{max}$ is then varied such that the computed solution fulfills
$\psi(s_\mathrm{max})\ll 1$ and matching the numerical and asymptotic solution.

For comparison we also give the approximate asymptotic analytical solution 
for the energy of the free membrane as calculated in \cite{foret2014}:
in the limit of a tense membrane ($R\gg \lambda$) one has 
\begin{align}
&\frac{E_\mathrm{Foret}^ \mathrm{tense}}{\kappa}
=4 \pi \bigg \{ 
\frac{4R}{\lambda} \sqrt{x(1-x)}(1-\sqrt{1-x})
\nonumber\\
&-x -2 \ln  \left ( \frac{1+\sqrt{1-x}}{2} 
\right ) -2 (1-\sqrt{1-x})^2 
\bigg \},
\label{eq:Foret_tense}
\end{align}
and for a loose membrane ($R\ll \lambda$)
\begin{align}
&\frac{E_\mathrm{Foret}^ \mathrm{loose}}{\kappa}=4 \pi 
\bigg( \frac{2R}{\lambda} x(1-x) \bigg)^2 \times
\nonumber\\
&\times
\bigg \{
- \gamma + \frac{x}{2(1-x)}-\ln \bigg( \frac{R}{\lambda}\sqrt{x(1-x)} (1-x) \bigg)
\bigg \},
\label{eq:Foret_loose}
\end{align}
where $x=(1-\cos  \theta)/2$ 
and the Euler constant $\gamma \approx 0.577$.

\section{Shape equations and energy of the free membrane for 
cylindrical particles with parallel orientation}
\label{appendix:free_membrane_parallel_cylinder}
For the parallel spherical particle the energy of the free membrane 
compared to the flat case can be written as 
\begin{equation}\label{eq:E_tot_free_cy_parallel}
\frac{E_\mathrm{free}^\mathrm{tot}}{\kappa} =
\int_0^\infty  \left (  \frac{ \dot{\psi} ^2 }{2}
+ \frac{ 1-\cos \psi }{\lambda^2}\right ) \, L \, \mathrm{d} s\,.
\end{equation}

where we use a similar parametrisation and the same geometric
relations as for the sphere
\begin{align} \label{eq:geometric_relations}
\dot{r}-\cos \psi&=0, \nonumber \\
\dot{z}+\sin \psi&= 0\,.
\end{align}

We construct a Lagrangian 

\begin{equation}\label{eq:Lagranian}
\mathcal{L} =
\dot{\psi} ^2 \frac{L}{2}
 + \frac{ L}{\lambda^2} (1-\cos \psi) \, ,
\end{equation}

and find the Euler-Lagrange equations to equal

\begin{align}
L \ddot{\psi} - \frac{L}{\lambda^2}  \sin \psi =0
\nonumber \\
\end{align}

From Eq.~(\ref{eq:Lagranian}) we construct a Hamiltonian by 

\begin{equation}\label{eq:Hamiltonian}
\mathcal{H} =\dot{r} \partial_{\dot{r}} \mathcal{L}
+\dot{\psi} \partial_{\dot{\psi}} \mathcal{L} -\mathcal{L} =
\dot{\psi} ^2 \frac{L}{2}
 - \frac{ L}{\lambda^2} (1- \cos \psi)\,.
\end{equation}

Since $\mathcal{L}$ does not depend on $s$, $\mathcal{H}$ is conserved.
Since the membrane is asymptotically flat $\mathcal{H}(s\rightarrow \infty)=0$ 
and thus $\mathcal{H}=0$ \cite{foret2018mechanosensitivity}.
Then, 

\begin{equation} \label{eq:psi_dot}
\dot{\psi}  =  \pm \sqrt{\frac{ 2}{\lambda^2} (1- \cos \psi)}\,.
\end{equation}

Using Eq.~(\ref{eq:psi_dot}) in Eq.~(\ref{eq:E_tot_free_cy_parallel})
we find

\begin{align}
\frac{E_\mathrm{free}^\mathrm{tot}}{\kappa} &=
2 \int_0^\infty  \left ( \frac{2}{\lambda^2} (1-\cos \psi )
\right ) \, L \, \mathrm{d} s\, , \nonumber \\
&= 2 L \sqrt{\frac{ 2}{\lambda^2}} \int_0^\theta \,
\sqrt{ (1- \cos \psi)}
\mathrm{d} \psi
\,, \nonumber \\
&= \frac{8 L}{\lambda} \left (1-\cos \frac{\theta}{2} \right ) 
\, ,
\end{align}
where the factor 2 correspond to the left and right side of the membrane
relative to the cylindrical particle.
The same calculation and result can be found in \cite{mkrtchyan2010adhesion}.


%

\end{document}